\documentclass[aps,twocolumn,prb,amsmath,amssymb,showpacs,superscriptaddress]{revtex4}

\usepackage{graphicx,epsfig}

\newcommand{\re}{\mathrm{Re\,}}
\newcommand{\im}{\mathrm{Im\,}}
\newcommand{\ei}{\mathrm{Ei}}
\newcommand{\sgn}{\mathrm{sgn}}

\begin{document}
 
\title{Screening of a Luttinger liquid wire by a scanning tunneling microscope tip:\\
I. Spectral properties}

\author{Marine Guigou}
\author{Thierry Martin}
\author{Adeline Cr\'epieux}

\affiliation{Centre de Physique Th\'eorique, Universit\'e de la M\'editerran\'ee, 163 avenue de Luminy, 13288 Marseille, France}

\date{\today}


\begin{abstract}
The screening effect due to a scanning tunneling microscope tip which is placed in the vicinity of an interacting quantum wire is considered. With the help of a bosonization procedure, we are able to determine non perturbatively the Green's functions of the quantum wire in the presence of both electrostatic screening by the tip and Coulomb interactions in the wire. 
In our approach we justify that the working Hamiltonian of the whole system is quadratic when $K_c>1/2$ and can be solved by integration over the degrees of freedom of the tip. Once the Green's functions are known, we calculate the spectral properties. We show that the spectral function, as well as the tunnel density of states, is affected by the screening and that the local density of states strongly deviates from its unscreened value when the tip gets close to the wire. Moreover, we observe that the spatial extension of the deviation of the local density of states is related to both the Coulomb interactions parameter and the screening strength.
\end{abstract}

\maketitle


\section{Introduction}

The scanning tunneling microscope (STM) is a useful tool to study surfaces at the atomic level. It allows to probe the local density of states of the material through the tunneling current \cite{binnig}, but it can also be used to move atoms\cite{li,dujardin} or a nanoscale object on a surface, for example a carbon nanotube\cite{gao}. The proximity of an electrical charged STM tip is not without consequences on the electronic properties of the nano-object or of the surface\cite{tromp,klitsner,pelz,mcellistrem}. Ab-initio approaches treating the tip plus the sample as a single system and including the tip-sample interaction have been developed\cite{cho,diventra}. However, most of these studies require the use of 
approximate schemes (such as density functional theory). 

The aim of our work is to study the effect of a STM tip on a one-dimensional system where Coulomb interactions are strong -- a Luttinger liquid quantum wire -- with potential applications to carbon nanotube experiments. 

One way to model the effect of the STM tip on the electronic structure of the quantum wire is to consider the tip as an impurity. 
The presence of a charged impurity near or in a quantum wire leads to screening effects. While the Thomas-Fermi approximation\cite{thomas,fermi} can be used, it is restricted to a gas of free electrons, and therefore does not apply
for a 1D strongly correlated system. 
Screening in an interacting wire can be treated by bosonization. It has been shown that the electroneutrality is insured when charges induced on nearby gates are taken into account\cite{egger2}. A similar study was performed for metallic carbon nanotubes\cite{sasaki}. The explicit screening by external gates has also been considered\cite{gonzalez}. In addition, it has been shown that a metallic carbon nanotube can screen a charged impurity, while a semiconducting nanotube cannot\cite{lin}.

The effect of a single impurity in a quantum wire has been widely studied\cite{egger1,luther,giamarchi,safi2,dolcini,schuricht}. But most of these studies are limited to perturbative calculations because of the $\cos(\sqrt{2\pi}\theta_c)$ term that appears due to backscattering by the impurity.

Another consequence of the presence of an impurity is the appearance of Friedel oscillations. Such Friedel oscillations have been considered for impurity in quantum wire\cite{urban,eggert,egger3} and the specificity for one-dimensional systems, in comparison to 2D or 3D systems, has been studied\cite{chaplik}. Exact calculations\cite{leclair} can be performed in the case where the Coulomb interactions parameter $K_c=1/2$.

Here we model the STM tip + wire system with the help of the Tomonaga-Luttinger liquid theory\cite{tomonaga,luttinger} which allows to take both the interactions between electrons in the wire and the screening by the tip into account. The electrostatic interaction between the tip and the wire depends on the densities of the two sub-systems.

In the wire, the density operator contains a slowly varying part as well as an oscillatory component which gives rise to backscattering. The latter contribution can be argued not to be relevant, which means that we are able to obtain non perturbative results for screening from the slowly varying 
contribution. The full Hamiltonian is quadratic even in the presence of the electrostatic interaction, but the screening interaction breaks translational invariance in the wire. 
The strategy of our work is to use this particularity in order to determine the exact Green's function of the wire in the presence of the tip and next, and to calculate the spectral properties and the transport properties of the whole system.

The spectral function is a useful tool in order to probe the electronic properties of a device, for example, the photoemission and inverse photoemission spectra can be interpreted to check the Luttinger liquid behavior of a quantum wire\cite{dardel}. It is well known that the weak dimensionality of a quantum wire or a carbon nanotube induces the appearance of the collectivization of excitations, with the respect of the spin and charge modes. This separation in terms of holons and spinons is directly viewed in the spectral function by the presence of several peaks which match to the spin-charge separation\cite{voit,meden}. One of the questions we want to address in this work is the effect of screening on the spectral function and on the density of states: is the spin-charge separation in the interacting wire affected by the presence of the tip? More importantly, is the power law in the density of states modified by the presence of the tip?

The article is organized as follows. In Sec.~II, we present the system we have considered and the model. In Sec.~III, we present the very general derivation of the Dyson equations associated to the wire + STM system in the presence of screening by the electrostatic potential. In Sec.~IV, general expressions for the spectral properties are given. Next, in Sec.~V, we apply our results to a specific screening potential and we give the explicit expressions of the Green's functions, the spectral function and the local density of states. We discuss their behaviors as a function of the position along the wire, the value of the Coulomb interaction parameter and the various energy scales/frequencies. We conclude in Sec.~VI. Technical details are given in appendices. The calculation of transport properties such as current and finite-frequency noise will be presented in a separate paper\cite{guigou}.


\section{Model}

\begin{figure}[ht]
\begin{center}
\includegraphics[width=5cm]{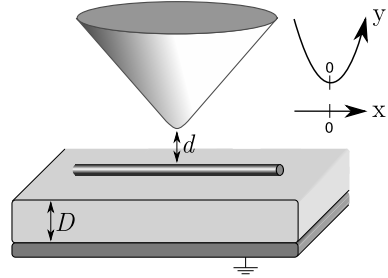}
\caption{Representation of the system. $d$ is the distance between the STM tip and the quantum wire and $D$ is the length which separates the wire and the gate.\label{system}}
\end{center}
\end{figure}

We consider the following system: a STM tip, which can be used as a probe as well as an electron injector, is placed in the vicinity 
of a metallic quantum wire (see Fig.~\ref{system}). The proximity between these two sub-systems induces changes in the electronic properties of the quantum wire, so that, the total Hamiltonian is composed by three parts $H=H_W+H_T+H_{Sc}$. The first part describes the quantum wire as an inhomogeneous non-chiral Luttinger liquid with spin degree of freedom:
\begin{eqnarray}
H_{W}&=&\sum_{j=c,s}\int_{-\infty}^{+\infty}dx \frac{v_{j}(x)}{2}\Big[K_{j}(x)\big(\partial_x\phi_{j}(x)\big)^2\nonumber\\
&+&K_{j}^{-1}(x)\big(\partial_x\theta_{j}(x)\big)^2\Big]~,
\label{H_W}
\end{eqnarray}

where $\theta_j$ and $\phi_j$ are the non-chiral bosonic fields which satisfy the commutation relation: $\big[\phi_{j}(x),\theta_{j'}(x')\big]_-=\big(i/2\big)\delta_{jj'}\sgn(x-x')$. 
The parameter $K_j$ is, in general, a function of the position $x$ along the quantum wire and is related to the strength of Coulomb interactions\cite{paradintcoul,safi3}. Its value depends on the sector: $j=c$ corresponds to the charge sector and $j=s$ corresponds to the spin sector. Because of $SU(2)$ invariant spin interactions, $K_s=1$ at any position. Inside the wire, we have $K_c=(1+2U_0/(v_F \pi))^{-1/2}<1$ since we assume repulsive interaction in the one-dimensional system. $U_0$ is the zero momentum Fourier transform of the interacting potential between the electrons. Outside the wire, we have $K_c=1$ since the electrodes are normal metals with no Coulomb interactions. The sector velocities are defined by $v_j(x)=v_F/K_j(x)$ where $v_F$ is the Fermi velocity of electrons in the wire.

The STM tip is assumed to be a normal metal but, for convenience\cite{STM}, it is better described with the help of  Luttinger liquid theory. The Hamiltonian encoding the information about this chiral liquid with infinite length is
\begin{eqnarray}
H_{T}&=&\frac{u_F}{4\pi}\sum_{\sigma'=\pm 1}\int_{-\infty}^{+\infty}dy~\big(\partial_y\varphi_{\sigma'}(y)\big)^2~,
\label{H_T}
\end{eqnarray}

with $\varphi_{\sigma'}$ is the chiral bosonic field of the STM tip, $u_F$ is the Fermi velocity of electrons in the tip and $y$ refers to the position with respect to the tip (see Fig~\ref{system}). Since the tip is a normal metal, its Coulomb interaction parameter is equal to one and there is no renormalisation of the Fermi velocity. The spin index $\sigma'$ allows us to consider a magnetic or non magnetic tip.\\
As we consider the possible screening effects between two sub-systems, the interaction describing this proximity  appears through the screening Hamiltonian:
\begin{eqnarray}
H_{Sc}=\int_{-\infty}^{+\infty} dx~ \int_{-\infty}^{+\infty} dy~\rho_{W}(x)W(x,y)\rho_{T}(y)~,
\label{H_Sc}
\end{eqnarray}

where we identify $W$ as the electrostatic potential coupling the densities and resulting from the electrostatic force caused by the proximity between the STM tip and the quantum wire. The bosonized form of the density operator of the quantum wire is defined by
\begin{eqnarray}
\rho_{W}(x)&=&\sum_{r,\sigma}\Big[\psi^{\dag}_{r\sigma}(x)\psi_{r\sigma}(x)\nonumber\\
&~&~~~+e^{-2irk_Fx}\psi^{\dag}_{r\sigma}(x)\psi_{-r\sigma}(x)\Big]~,
\end{eqnarray}

where $\psi_{r\sigma}$ is the fermionic field which describe the propagation of excitations with spin $\sigma$ in the wire and is decomposed in term of right-mover ($r=+$) and left-mover ($r=-$) which can be expressed through the bosonic operators\cite{haldane} $\theta_j$ and $\phi_j$:
\begin{eqnarray}
\psi_{r\sigma}(x)&=&\frac{f_{r\sigma}}{\sqrt{2\pi a}}e^{irk_Fx+i\sqrt{\frac{\pi}{2}}\sum_{\sigma}h_{\sigma}(j)\big[\phi_j(x)+r\theta_j(x)\big]}~,\nonumber\\
\label{psi}
\end{eqnarray}

where $f_{r\sigma}$ is a Majorana fermion, which insures the correct anti-commutation relation for the fermionic fields, $k_F$ is the Fermi wave vector and, $a$ is the ultraviolet cut-off. The coefficient $h_\sigma(j)$ is equal to $1$ when $j=c$ and to $\sigma$ when $j=s$. Using this definition, 
\begin{eqnarray}
\label{slow_fast}
\rho_{W}(x)&=&\sqrt{\frac{2}{\pi}}\partial_x \theta_{\mathrm{c}}(x)+\rho_{2k_{\mathrm{F}}}(x),
\end{eqnarray}

where the first (second) term corresponds to the slowly (rapidly) varying part of the wire density operator, and the rapidly oscillating component reads:
\begin{eqnarray}
\label{CDW}
\rho_{2k_{\mathrm{F}}}(x)&=&\frac{2k_{\mathrm{F}}}{\pi}\cos\big(2k_{\mathrm{F}}+\sqrt{2\pi}\theta_{\mathrm{c}}(x)\big)\nonumber\\
&&\times \cos\big(\sqrt{2\pi}\theta_{\mathrm{s}}(x)\big)~.
\end{eqnarray} 

This is precisely the contribution which enters the expression of the backscattering potential of a localized impurity in Ref.~\onlinecite{kane}. 

The bosonised version of the tip density reads:
\begin{eqnarray}
\rho_{T}(y)&=&\frac{1}{2\pi}\sum_{\sigma'}\partial_y\varphi_{\sigma'}(y)~.
\end{eqnarray}

Notice that if one wanted to model a carbon nanotube instead of a single channel quantum wire, it would be sufficient to insert in the Hamiltonian two additional degrees of freedom (denoted by the index $\delta$ in Ref.~\onlinecite{STM}): we would then have four sectors\cite{egger4} (total charge, relative charge, total spin and relative spin) instead of the two sectors (charge and spin) which we have here.


\section{Dyson equations}

The spectral properties can be calculated from the Green's function of the global system. 
Here we show that -- apart from irrelevant terms which can in principle be evaluated 
safely within the context of perturbation theory -- it is possible to obtain non perturbatively 
the Green's function of the total system, 
including electrostatic screening. 

The first step is to write the partition function in the Matsubara formalism\cite{QFT}:
\begin{eqnarray}
Z[\varphi,\phi,\theta]&=&\int {\mathcal D}\varphi_{\uparrow}{\mathcal D}\varphi_{\downarrow}{\mathcal D}\phi_c{\mathcal D}\phi_s{\mathcal D}\theta_c{\mathcal D}\theta_s e^{-\int d\tau (L-L_{aux})}~,\nonumber\\
\end{eqnarray}

where $L=L_W+L_T+L_{Sc}$ is the total Lagrangian of the system and $L_{aux}$ contains auxiliary fields $\eta_{\theta_j}$, $\eta_{\phi_j}$ and $\eta_{\varphi_{\sigma'}}$ needed to extract the Green's functions, such as for example ${\bf G}_j^{\theta\theta}(x,\tau;x',\tau')=\langle T_{\tau}\{\theta_j(x,\tau)\theta_j(x',\tau')\}\rangle$, through the relation:
\begin{eqnarray}
\mathbf{G}_j^{\theta\theta}(x,\tau;x',\tau')
=\frac{1}{Z}\frac{\partial^2Z}{\partial\eta_{\theta_j}(x,\tau)\partial\eta_{\theta_j}(x',\tau')}~,
\end{eqnarray}

where $T_{\tau}$ is the time ordering operator.

According to Eq.~(\ref{slow_fast}), 
the total Hamiltonian is quadratic in the bosonic fields, except for the presence of the $2k_F$ contribution of the wire
density in the screening interaction. Let us discuss why we can neglect this contribution. 
If one integrates out the tip degrees of freedom first, the screening interaction generates
an effective, retarded backscattering interaction which is a convolution
of $\rho_{2k_{\mathrm{F}}}(x)$, $\rho_{2k_{\mathrm{F}}}(x')$, derivatives of the screening potentials $\partial_x\partial_y W(x,y)$, $\partial_{x'}\partial_{y'} W(x',y')$
and the tip Green's function.
As is explicitly shown below (Sec.~\ref{local_section}), the screening potential varies slowly on the scale of the Fermi wave length: 
it has a power law decay in $x$. A qualitative argument to neglect this retarded backscattering interaction is 
to state that rapid oscillations over the effective range of $W(x,y)$ lead to its vanishing. 
But one can also go further and examine the relevance of this interaction. 
Sure, the screening potential has a power law decay, yet this decay is fast enough 
-- $\int dx \partial_x\partial_y W(x,y)$ converges -- to treat the derivative of the screening potential
as a delta function potential at $x=0$. Therefore, integrating out the tip leads to a backscattering
interaction which is proportional to $(\rho_{2k_{\mathrm{F}}}(0))^2$.
Now, it has been shown\cite{kane} that the renormalization group equations for the amplitude of a backscattering potential 
$v_{n_c,n_s}\cos\big(\sqrt{2\pi}n_c\theta_{\mathrm{c}}(0)\big)\cos\big(\sqrt{2\pi}n_s\theta_{\mathrm{s}}(0)\big)$
($n_c$, $n_s$ integers) read:
\begin{eqnarray}
\label{scaling}
\frac{\partial v_{n_c,n_s}}{\partial l}=\left(1-\frac{n_c^2}{2}K_c-\frac{n_s^2}{2}K_s\right) v_{n_c,n_s}~.
\end{eqnarray}

In particular, this is the well known result
which states that single electron backscattering ($n_c=n_s=1$) becomes relevant for 
repulsive interaction $K_c<1$ (here $K_s=1$).
In contrast, the backscattering contribution which is generated by the screening interaction
has {\it two} powers of $\rho_{2k_{\mathrm{F}}}(x)$. This contains contributions with
$\{n_c=n_s=2\}$, $\{n_c=2$, $n_s=0\}$, and $\{n_c=0$, $n_s=2\}$. 
By simple power counting, we estimate that 
the backscattering term is irrelevant in the renormalization group sense as long as $K_c>1/2$.
While it is true that some carbon nanotubes (in an insulating environment) are expected to have an interaction parameter
$K_c$ as low as $0.2-0.3$\cite{egger4}, the presence of the nearby substrate/gate provides additional screening which 
tends to enhance $K_c$ and to bring it closer to the non interacting value $1$.  

For this reason, in the present calculation, only the slow component of the wire 
density in Eq.~(\ref{slow_fast}) is kept, so that
the Hamiltonian/Lagrangian are quadratic in the bosonic fields. 
By integrating out the degree of freedom of the tip (wire), we are thus able to extract the Dyson equation of the bosonic Green's functions for the wire (tip). The details of the calculation are presented in Appendix A. 

Once we integrate out the degrees of freedom of the tip, we obtain the following Dyson equation for the full Green's function ${\bf G}_j^{\theta\theta}$ of the wire in the presence of screening:
\begin{widetext}
\begin{eqnarray}\label{DEG}
\mathbf{G}^{\theta\theta}_j(x,\tau;x',\tau')&=&G^{\theta\theta}_j(x,\tau;x',\tau')+\delta_{jc}\sum_{\sigma'}\int d\tau_1\int d\tau_2\int_{-\infty}^{+\infty}dx_1\int_{-\infty}^{+\infty}dx_2\int_{-\infty}^{+\infty}dy_1\int_{-\infty}^{+\infty}dy_2\nonumber\\
&\times& G^{\theta\theta}_j(x,\tau;x_1,\tau_1)G^{-1}_{Sc}(x_1,y_1)G_{\sigma'}^{\varphi\varphi}(y_1,\tau_1;y_2,\tau_2)G^{-1}_{Sc}(x_2,y_2)\mathbf{G}^{\theta\theta}_j(x_2,\tau_2;x',\tau')~,
\end{eqnarray}
\end{widetext}

where $G^{\theta\theta}_j$ is the bare Green's function of the wire (i.e., in the absence of screening), $G^{-1}_{Sc}(x,y)=\partial_x\partial_yW(x,y)/(\pi\sqrt{2\pi})$ and $G_{\sigma'}^{\varphi\varphi}$ is the bare Green's function of the tip. It is important to note that the screening effects which we consider here play a role only in the charge sector, which means that all the Green's functions in the spin sector ($j=s$) are the same as the bare ones. In a similar way, we obtain the Dyson equations associated to the Green's functions of the wire $\mathbf{G}^{\phi\phi}_j$, $\mathbf{G}^{\phi\theta}_j$ and $\mathbf{G}^{\theta\phi}_j$.

When we integrate over the degrees of freedom of the wire, we obtain the Dyson equation for the Green's function of the tip $\mathbf{G}_{\sigma'}^{\varphi\varphi}$ in the presence of screening:
\begin{widetext}
\begin{eqnarray}\label{dyson3}
\mathbf{G}_{\sigma'}^{\varphi\varphi}(y,\tau;y',\tau')&=&G_{\sigma'}^{\varphi\varphi}(y,\tau;y',\tau')+\sum_{j}\delta_{jc}\int d\tau_1\int d\tau_2 \int_{-\infty}^{+\infty}dx_1\int_{-\infty}^{+\infty}dx_2\int_{-\infty}^{+ \infty}dy_1\int_{-\infty}^{+\infty}dy_2\nonumber\\
&\times&G_{\sigma'}^{\varphi\varphi}(y,\tau;y_1,\tau_1)G^{-1}_{Sc}(x_1,y_1)G_j^{\theta\theta}(x_1,\tau_1;x_2,\tau_2)G^{-1}_{Sc}(x_2,y_2)\mathbf{G}_{\sigma'}^{\varphi\varphi}(y_2,\tau_2;y',\tau')~.
\end{eqnarray}
\end{widetext}


At this stage, the Dyson equations which we have obtained are very general since we do not need to specify neither the expressions of the bare Green's functions, nor the nature of the potential $W$ describing the interaction between the quantum wire and the STM tip. They can be applied to an inhomogeneous wire (i.e., with a Coulomb interactions parameter $K_c$ which depends of the position) as well as to a homogenous wire. They can also be applied to a system with more degrees of freedom such as a carbon nanotube. In that case, as explained in Sec.~II, it is sufficient to add an index $\delta$ to the Green's functions of the wire.


\section{Spectral properties}

In this section, we give the general expressions of the spectral function and of the local density of states as a function of the bosonic Green's functions in the presence of screening. 

\subsection{Spectral function definition}

For a homogenous wire, the spectral function associated to the spin $\sigma$ and the chirality $r$ is connected to the retarded Green's function by
\begin{eqnarray}\label{def_SF}
A_{r,\sigma}(k,\Omega)&\equiv&-\frac{1}{\pi}\im[{\cal G}^R_{r,\sigma}(k,\Omega)]~,
\end{eqnarray}

where ${\cal G}^R_{r,\sigma}(k,\Omega)$ is the double Fourier transform of the retarded Green's function ${\cal G}^R_{r,\sigma}(\tilde{x},\tilde{t})$ defined as the average of the anti-commutator of the fermionic fields:
\begin{eqnarray}
{\cal G}^R_{r,\sigma}(\tilde{x},\tilde{t})
&\equiv&-i\Theta(t-t')\langle[\psi_{r\sigma}(x,t),\psi_{r\sigma}^{\dag}(x',t')]_+\rangle~,\nonumber\\
\end{eqnarray}

with $\Theta$, the Heaviside function. The Fourier transforms are performed over the argument $\tilde{t}=t-t'$ using the time translational invariance and over $\tilde{x}=x-x'$ using the spatial translational invariance. 

The screening by the STM tip located at $x=0$ (see Fig.~\ref{system}) breaks the spatial translation invariance and thus, contrary to the homogeneous case, the fermionic Green's function will depend not only on the distance $x-x'$, but both on the positions $x$ and $x'$. After performing the change of variables: $\tilde{x}=x'-x$ and $\tilde{X}=(x+x')/2$, we will see in Sec.~V that the fermionic Green's function does not have a strong dependence on $\tilde{X}$ and in a first approximation, we will define a spectral function as in Eq.~(\ref{def_SF}) even in the presence of screening. However, when writing of the Green's functions, we keep the $x$ and $x'$ dependences for the sake of clarity.

With the help of the relation ${\cal G}^R_{r,\sigma}-{\cal G}^A_{r,\sigma}={\cal G}^{-+}_{r,\sigma}-{\cal G}^{+-}_{r,\sigma}$, we can express the spectral function in term of the Keldysh Green's functions:
\begin{eqnarray}
A_{r,\sigma}(k,\Omega)&=&-\frac{1}{2\pi}\im\Big[{\cal G}^{-+}_{r,\sigma}(k,\Omega)-{\cal G}^{+-}_{r,\sigma}(k,\Omega)\Big]~.
\end{eqnarray}

Next, using Eq.~(\ref{psi}), we write the Fourier transforms of the fermionic Keldysh Green's function in terms of the bosonic Keldysh Green's functions extracted from the Dyson equation derived in Sec.~III, and we determine their Keldysh components \cite{keldysh} by means of analytic continuations $\tau\rightarrow it+\tau_0$ and $\tau'\rightarrow it'+\tau_0$. After some algebra, we obtain:
\begin{eqnarray}\label{SF}
&&{\cal G}^{-+}_{r,\sigma}(x,x',\tilde{t})=-i\frac{e^{irk_F(x-x')}}{2\pi a}\nonumber\\
&&\times e^{\frac{\pi}{2}\sum_{j}\mathbf{G}_{j}^{\phi\phi,-+}(x,x',\tilde{t})+\frac{\pi}{2}\sum_{j}\mathbf{G}_{j}^{\theta\theta,-+}(x,x',\tilde{t})}\nonumber\\
&&\times e^{\frac{\pi}{2}\sum_{j}r\mathbf{G}_{j}^{\phi\theta,-+}(x,x',\tilde{t})+\frac{\pi}{2}\sum_{j}r\mathbf{G}_{j}^{\theta\phi,-+}(x,x',\tilde{t})}~,
\end{eqnarray}

where ${\bf G}_j^{\theta\theta,\eta\eta'}$ is the Keldysh bosonic Green's function in the presence of screening defined as:
\begin{eqnarray}\label{Definition_Green}
&&{\bf G}_j^{\theta\theta,\eta\eta'}(x,x',\tilde{t})\equiv{\bf G}_j^{\theta\theta,\eta\eta'}(x,t^\eta;x',t'^{\eta'})\nonumber\\
&=&\langle\theta_j(x,t^\eta)\theta_j(x',t'^{\eta'})\rangle-\frac{1}{2}\langle\theta_j(x,t^\eta)\theta_j(x,t^\eta)\rangle\nonumber\\
&&-\frac{1}{2}\langle\theta_j(x',t'^{\eta'})\theta_j(x',t'^{\eta'})\rangle~,
\end{eqnarray}

and similarly for ${\bf G}_j^{\theta\phi,\eta\eta'}$, ${\bf G}_j^{\phi\theta,\eta\eta'}$ and ${\bf G}_j^{\phi\phi,\eta\eta'}$. $\eta=\pm$ and $\eta'=\pm$ are the Keldysh indices. 

\subsection{Definition of the density of states}

The density of states in the wire at position $x$ associated to the spin $\sigma$ and the chirality $r$ is defined by:
\begin{eqnarray}
\rho_{r,\sigma}(x,\Omega)&\equiv&-\frac{1}{\pi}\im[{\cal G}^R_{r,\sigma}(x,x,\Omega)]~,
\end{eqnarray}

where 
\begin{eqnarray}
{\cal G}^R_{r,\sigma}(x,x,\Omega)=~\int_{-\infty}^{+\infty}d\tilde{t}e^{i\Omega\tilde{t}}{\cal G}^R_{r,\sigma}(x,x,\tilde{t}).
\end{eqnarray}

When the bosonic Green's functions are injected in the expression of the fermionic Green's function, one obtains for the density of states
\begin{widetext}
\begin{eqnarray}\label{DOS}
\rho_{r,\sigma}(x,\Omega)&=&\frac{1}{4\pi^2 a}\re\Bigg[\int_{-\infty}^{+\infty}d\tilde{t} e^{i\Omega \tilde{t}}\Big(e^{\frac{\pi}{2}\sum_{j}\mathbf{G}_{j}^{\phi\phi,-+}(x,x,\tilde{t})+\frac{\pi}{2}\sum_{j}\mathbf{G}_{j}^{\theta\theta,-+}(x,x,\tilde{t})+\frac{\pi}{2}\sum_{j}r\mathbf{G}_{j}^{\phi\theta,-+}(x,x,\tilde{t})+\frac{\pi}{2}\sum_{j}r\mathbf{G}_{j}^{\theta\phi,-+}(x,x,\tilde{t})}\nonumber\\
&&+e^{\frac{\pi}{2}\sum_{j}\mathbf{G}_{j}^{\phi\phi,+-}(x,x,\tilde{t})+\frac{\pi}{2}\sum_{j}\mathbf{G}_{j}^{\theta\theta,+-}(x,x,\tilde{t})+\frac{\pi}{2}\sum_{j}r\mathbf{G}_{j}^{\phi\theta,+-}(x,x,\tilde{t})+\frac{\pi}{2}\sum_{j}r\mathbf{G}_{j}^{\theta\phi,+-}(x,x,\tilde{t})}\Big)\Bigg]~.
\end{eqnarray}
\end{widetext}

In the next section, we present the calculation of these quantities for an infinite length wire and a local electrostatic potential double derivative.


\section{Application to a local electrostatic potential double derivative $\partial_x\partial_yW(x,y)$}
\label{local_section}

The electrostatic potential induced by the STM tip depends of the geometry of the system, in particular it depends on the distance between the tip and the wire, on the distance between the tip and the gate and on the shape of the tip\cite{model_STM1,model_STM2}. We assume that the tip acts as a local charge. With the help of the image charge method, one can determine the precise form of the potential\cite{ICM1,ICM2} for the system depicted in Fig.~$\ref{system}$, we obtain:
\begin{eqnarray}\label{potential}
W(x,y)&=&\frac{e^2}{4\pi\varepsilon_0}\left[\frac{1}{\sqrt{x^2+(y+d)^2}}\right.\nonumber\\
&&\left.-\frac{1}{\sqrt{x^2+(y+d+2D)^2}}\right]~,
\end{eqnarray}

where $d$ is the distance between the tip and the wire and $D$ is the distance between the wire and the gate. The quantity which appears in the Dyson equation is the double derivative of the potential $\partial_x\partial_yW(x,y)$. From Eq.~(\ref{potential}), it can be shown that the double derivative behaves as $x^{-4}$ for large $x$ and as $y^{-4}$ for large $y$. These fast decreases with $x$ and $y$ express the fact that the main action of the potential is centered on positions $y=0$ for the tip and $x=0$ for the wire. As a consequence, we assume a local double derivative for the potential and use for the calculations of the Green's functions the expression
\begin{eqnarray}
\partial_x\partial_yW(x,y)&=&W_0\delta(x)\delta(y)~,
\label{derive_potential}
\end{eqnarray}

where $W_0\equiv e^2/(4\pi\varepsilon_0 d)$ is the screening potential strength: it increases when the distance $d$ between the tip and the wire decreases. 

With this specific form of the electrostatic potential and for an infinite length wire, we are able to solve the Dyson equations of Sec.~III. Let us first report Eq.~(\ref{derive_potential}) in the Dyson equation of the bosonic Green's function $\mathbf{G}^{\theta\theta}_c$ given by Eq.~(\ref{DEG}). Taking the Fourier transform $\mathbf{G}^{\theta\theta}_{c}(x,x',\omega)=\int\mathbf{G}^{\theta\theta}_{c}(x,x',\tilde{\tau})\exp(i\omega \tilde{\tau})d\tilde{\tau}$, one obtains in the Matsubara formalism 
\begin{eqnarray}\label{equat_green}
&&\mathbf{G}^{\theta\theta}_c(x,x',\omega)=G^{\theta\theta}_c(x,x',\omega)\nonumber\\
&&+\sum_{\sigma'}\frac{W_0^2}{2\pi^3}G^{\theta\theta}_c(x,0,\omega)G_{\sigma'}^{\varphi\varphi}(0,0,\omega)\mathbf{G}^{\theta\theta}_c(0,x',\omega)~.\nonumber\\
\end{eqnarray}

With the help of the bare Green's functions at zero temperature (given in Appendix B), Eq.~(\ref{equat_green}) can be solved exactly. We obtain:
\begin{eqnarray}
\mathbf{G}^{\theta\theta}_{c}(x,x',\omega)&=&G^{\theta\theta}_{c}(x,x',\omega)\nonumber\\
&+&\frac{K_c\omega_{Sc}}{2|\omega|(\omega^2-\omega_{Sc}^2)}e^{-K_c|\omega|\frac{|x|+|x'|}{v_F}}~,
\end{eqnarray}

where $\omega_{Sc}\equiv(W_0/\pi)\sqrt{K_c/2}$ is the frequency associated with the screening. Next, we perform an inverse Fourier transform: $\mathbf{G}^{\theta\theta}_{c}(x,x',\tilde{\tau})=\int\mathbf{G}^{\theta\theta}_{c}(x,x',\omega)\exp(-i\omega \tilde{\tau})d\omega/(2\pi)$. From this Matsubara Green's function, one can derive the Keldysh Green's functions (see Appendix C for details) by performing an analytic continuation $\tilde{\tau}\rightarrow i\tilde{t}+\tau_0$, where $\tau_0=a/v_F$, we obtain:
\begin{eqnarray}\label{green1}
&&\mathbf{G}^{\theta\theta,-+}_{c}(x,x',\tilde{t})=G^{\theta\theta,-+}_{c}(x,x',\tilde{t})\nonumber\\
&&+\frac{K_c}{8\pi}\sum_r\Bigg[2i\pi r\cosh\left(\omega_{Sc}\chi _r\right)+2\ln\left(\chi _r\right)\nonumber\\
&&-e^{\omega_{Sc}\chi _r}\ei\left(-\omega_{Sc}\chi _r\right)
-e^{-\omega_{Sc}\chi _r}\ei\left(\omega_{Sc}\chi _r\right)\Bigg]~,
\end{eqnarray}

where $\chi _r\equiv -r\tilde{t}-K_c(|x|+|x'|)/v_F+iar/v_F$ and $\ei$ is the exponential integral function. Notice that Eq.~(\ref{green1}) is valid whatever the amplitude of the screening is since no assumption has been made on the strength of the screening potential. We recall that for the spin sector ($j=s$), the Green's function is not affected by the screening: $\mathbf{G}^{\theta\theta,-+}_{s}(x,x',\tilde{t})=G^{\theta\theta,-+}_{s}(x,x',\tilde{t})$. The other Keldysh Green's functions are given in Appendix C. Because of the definition of Eq.~(\ref{Definition_Green}), the Green's functions that we have to use in the spectral function and the density of states calculations must be shifted by:
\begin{eqnarray}
&&\mathbf{G}^{\theta\theta,\eta\eta'}_{j}(x,x',\tilde{t})\rightarrow\mathbf{G}^{\theta\theta,\eta\eta'}_{j}(x,x',\tilde{t})\nonumber\\
&&-\frac{1}{2}\mathbf{G}^{\theta\theta,\eta\eta}_{j}(x,x,0)-\frac{1}{2}\mathbf{G}^{\theta\theta,\eta'\eta'}_{j}(x',x',0)~.
\end{eqnarray}

We remark that in the absence of screening (i.e., $\omega_{Sc}=0$), the full Green's functions reduce to the bare Green's functions $\mathbf{G}^{\theta\theta,\eta\eta'}_{j}(x,x',\tilde{t})=G^{\theta\theta,\eta\eta'}_{j}(x,x',\tilde{t})$ since\cite{gradshteyn} 
\begin{eqnarray}
\lim_{\alpha\rightarrow 0}\ei(\alpha)= C+\ln(\alpha)~,
\end{eqnarray}

where $C$ is the Euler-Mascheroni constant. \\

\subsection{Spectral function}

Reporting Eq.~(\ref{green1}) and the expressions of the other bosonic Green's functions, given in Appendix C, in Eq.~(\ref{SF}), we obtain the fermionic Green's function in the presence of screening:
\begin{widetext}
\begin{eqnarray}
&&{\cal G}^{-+}_{r,\sigma}(x,x',\tilde{t})=-i\frac{e^{irk_F(x-x')}}{2\pi a}e^{F_0^r(x,x',\tilde{t})+\sum_{r'}\frac{(K_c+rr'\sgn(x))(K_c-rr'\sgn(x'))}{16K_c}\left(F_{Sc}^{r'}(x,x',\tilde{t})-\frac{1}{2}F_{Sc}^{r'}(x,x,0)-\frac{1}{2}F_{Sc}^{r'}(x',x',0)\right)}~.
\end{eqnarray}
\end{widetext}

The function $F_0^r$ contains all the terms that are present in the exponential in absence of screening\cite{voit,meden}:
\begin{eqnarray}
&&F_0^r(x,x',\tilde{t})=-\frac{1}{2}\sum_j\ln\left(1+i\omega_c\tilde{t}-irK_j\frac{x-x'}{a}\right)\nonumber\\
&&-\sum_{r',j} \gamma_j \ln\left(1+i\omega_c\tilde{t}+ir'K_c\frac{x-x'}{a}\right)~,
\end{eqnarray}

where $\omega_c\equiv v_F/a$ is the frequency cut-off and $\gamma_c\equiv(K_c+K_c^{-1}-2)/8$. Since we look at a system in which the interactions are spin-independent, the parameter related to the spin sector, $\gamma_s$, is equal to zero. We remark that $F_0^r$ does depend only on the distance $(x-x')$ because it corresponds to the homogenous wire contribution for which translation invariance holds.

The function $F_{Sc}^{r'}$ corresponds to the screening part and it is given by:
\begin{eqnarray}\label{fct_screen}
&&F_{Sc}^{r'}(x,x',\tilde{t})=2i\pi r'\cosh\left(\omega_{Sc}\chi _{r'}\right)+2\ln\left(\chi _{r'}\right)\nonumber\\
&&-e^{\omega_{Sc}\chi _{r'}}\ei\left(-\omega_{Sc}\chi _{r'}\right)
-e^{-\omega_{Sc}\chi _{r'}}\ei\left(\omega_{Sc}\chi _{r'}\right)~.
\end{eqnarray}

\begin{figure}[ht]
\begin{center}
\includegraphics[width=4.2cm]{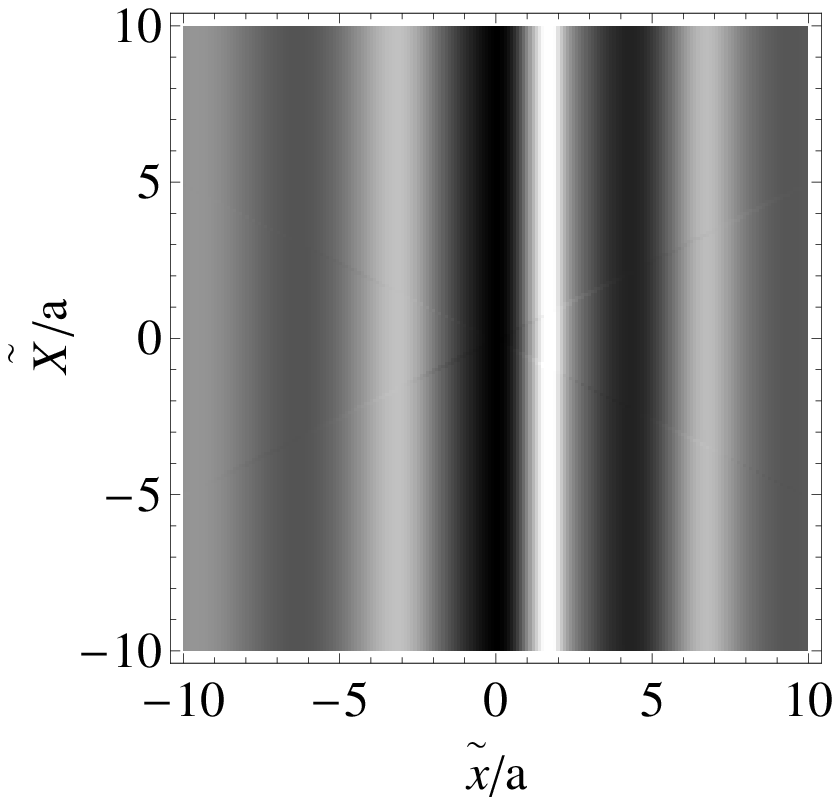}
\includegraphics[width=4.2cm]{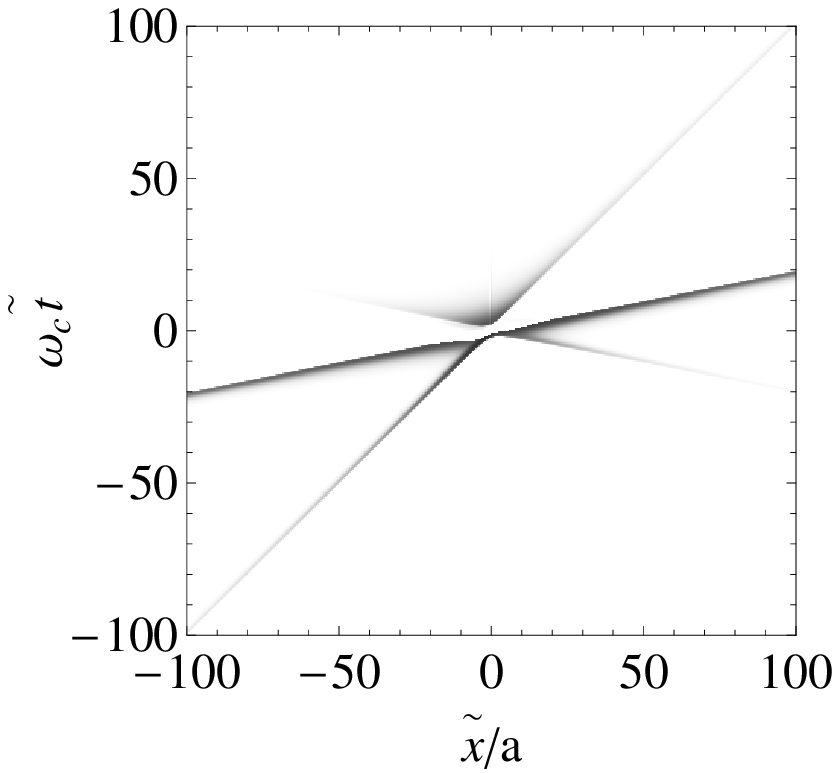}
\caption{Left panel: projected view of the real part of the fermionic Green's function ${\cal G}^{-+}_{+,\sigma}$ as a function of $\tilde{x}/a$ and $\tilde{X}/a$ for a fixed value of $\omega_c\tilde{t}$, at $K_c=0.2$ and $\omega_{Sc}/\omega_c=0.005$. Right panel: the same quantity is plotted as a function of $\tilde{x}/a$ and $\omega_c\tilde{t}$. On this second graphic, the $\tilde{X}$-dependence has been neglected. The imaginary part of ${\cal G}^{-+}_{+,\sigma}$ shows the same kind of behavior.}\label{figurePROJ}
\end{center}
\end{figure}

Since $\chi _{r'}\equiv -r'\tilde{t}-K_c(|x|+|x'|)/v_F+iar'/v_F$, we see that the fermionic Green's function depends both on $x$ and $x'$. We perform the change of variables: $\tilde{x}=(x-x')$ and $\tilde{X}=(x+x')/2$, and we study the variation of ${\cal G}^{-+}_{r,\sigma}(\tilde{x},\tilde{X},\tilde{t})$ with the distance $\tilde{x}$ and the barycenter $\tilde{X}$. We observe a very weak dependence with $\tilde{X}$ whereas we have a strong dependence with $\tilde{x}$ (see the left panel of Fig.~\ref{figurePROJ}): thus, we neglect the dependence with $\tilde{X}$ in a first approximation. As a consequence, even in the presence of screening, it is possible to define a single $k$-dependent spectral function:
\begin{eqnarray}\label{singleSF}
A_{r,\sigma}(k,\Omega)&=&-\frac{1}{2\pi}\im\Bigg[\int_{-\infty}^{+\infty}d\tilde{t}\int_{-\infty}^{+\infty}d\tilde{x}e^{i\Omega \tilde{t}-ik\tilde{x}}\nonumber\\
&&\times\left({\cal G}^{-+}_{r,\sigma}(\tilde{x},\tilde{t})-{\cal G}^{+-}_{r,\sigma}(\tilde{x},\tilde{t})\right)\Bigg]~,
\end{eqnarray}

where
\begin{eqnarray}
&&{\cal G}^{-+}_{r,\sigma}(\tilde{x},\tilde{t})=-i\frac{e^{irk_F\tilde{x}}}{2\pi a}\nonumber\\
&&\times\Bigg[e^{\tilde{F}_0^r(\tilde{x},\tilde{t})+\sum_{r'}\frac{(K_c+rr'\sgn(x))^2}{16K_c}\left(\tilde{F}_{Sc}^{r'}(\tilde{x},\tilde{t})-\tilde{F}_{Sc}^{r'}(\tilde{x},0)\right)}\Bigg]~,\nonumber\\
\end{eqnarray}

with
\begin{eqnarray}
&&\tilde{F}_0^r(\tilde{x},\tilde{t})=-\frac{1}{2}\sum_j\ln\left(1+i\omega_c\tilde{t}-irK_j\frac{\tilde{x}}{a}\right)\nonumber\\
&&-\sum_{r',j} \gamma_j \ln\left(1+i\omega_c\tilde{t}+ir'K_c\frac{\tilde{x}}{a}\right)~,
\end{eqnarray}

and,
\begin{eqnarray}
&&\tilde{F}^{r'}_{Sc}(\tilde{x},\tilde{t})=2i\pi r'\cosh\left(\omega_{Sc}\tilde{\chi} _{r'}\right)+2\ln\left(\tilde{\chi} _{r'}\right)\nonumber\\
&&-e^{\omega_{Sc}\tilde{\chi}_{r'}}\ei\left(-\omega_{Sc}\tilde{\chi}_{r'}\right)
-e^{-\omega_{Sc}\tilde{\chi}_{r'}}\ei\left(\omega_{Sc}\tilde{\chi}_{r'}\right)~,
\end{eqnarray}

where $\tilde{\chi}_{r'}\equiv -r'\tilde{t}-K_c|\tilde{x}|/v_F+iar'/v_F$. The real part of the fermionic Green's function ${\cal G}^{-+}_{+,\sigma}$ is plotted on the right panel of Fig.~\ref{figurePROJ}. We clearly see that the main contributions to the weight of the spectral function come from the lines for which $\omega_c\tilde{t}=\tilde{x}/a$ (i.e., $\tilde{x}=v_F\tilde{t}$), $\omega_c\tilde{t}=K_c\tilde{x}/a$ (i.e., $\tilde{x}=v_c\tilde{t}$), and $\omega_c\tilde{t}=-K_c\tilde{x}/a$ (i.e., $\tilde{x}=-v_c\tilde{t}$): these values cancel the imaginary parts in the logarithm arguments of $\tilde{F}_0^r$. It is then possible to reduce the double integral which appears in Eq.~(\ref{singleSF}) to a single one. The spectral function can thus be obtained from the three contributions:
\begin{eqnarray}
A_{1,r,\sigma}(k,\Omega)&&=-\frac{1}{2\pi}\im\Bigg[\int_{-\infty}^{+\infty}d\tilde{t}e^{i(\Omega-v_Fk)\tilde{t}}\nonumber\\
&&\times\left({\cal G}^{-+}_{r,\sigma}(v_F\tilde{t},\tilde{t})-{\cal G}^{+-}_{r,\sigma}(v_F\tilde{t},\tilde{t})\right)\Bigg]~,\\
A_{2,r,\sigma}(k,\Omega)&&=-\frac{1}{2\pi}\im\Bigg[\int_{-\infty}^{+\infty}d\tilde{t}e^{i(\Omega-v_ck)\tilde{t}}\nonumber\\
&&\times\left({\cal G}^{-+}_{r,\sigma}(v_c\tilde{t},\tilde{t})-{\cal G}^{+-}_{r,\sigma}(v_c\tilde{t},\tilde{t})\right)\Bigg]~,\\
A_{3,r,\sigma}(k,\Omega)&&=-\frac{1}{2\pi}\im\Bigg[\int_{-\infty}^{+\infty}d\tilde{t}e^{i(\Omega+v_ck)\tilde{t}}\nonumber\\
&&\times\left({\cal G}^{-+}_{r,\sigma}(-v_c\tilde{t},\tilde{t})-{\cal G}^{+-}_{r,\sigma}(-v_c\tilde{t},\tilde{t})\right)\Bigg]~.\nonumber\\
\end{eqnarray}

Using this simplification, we are able to calculate numerically the spectral function both in the absence and in the presence of screening. However, we must empathize that restrictions over the ranges of frequencies play an important role in this integration and must be carefully taken into account.

\begin{figure}[h]
\begin{center}
\includegraphics[width=8cm]{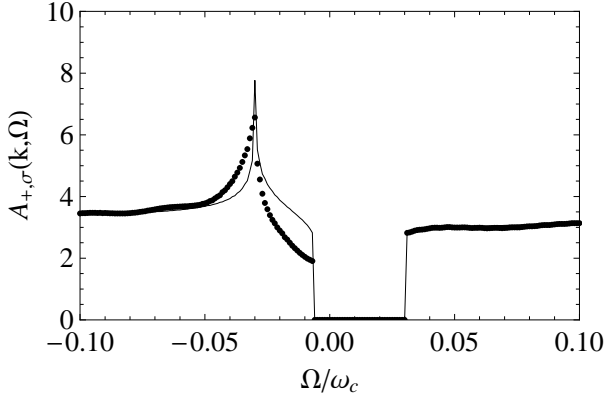}
\caption{Spectral function of the quantum wire for $K_c=0.2$, $a(k-k_F)=0.006$, in the absence of screening: $\omega_{Sc}/\omega_c=0$ (solid line), and in the presence of screening: $\omega_{Sc}/\omega_c=0.005$ (dots).
}\label{figureSPEC2}
\end{center}
\end{figure}

In Fig.~\ref{figureSPEC2}, we show the spectral function of the wire for strong Coulomb interactions ($K_c=0.2$). In the absence of screening (solid line), we recover the spectral function of an infinite interacting quantum wire that has been calculated analytically in Refs.~\onlinecite{meden} and \onlinecite{voit}. The main characteristics in this spectral function is the presence of singularities at $\Omega=\pm v_c(k-k_F)$, which correspond to charge excitations, and at $\Omega=-v_F(k-k_F)$, which corresponds to spin excitations. The fact that these singularities appear at different frequencies in the spectral function is the witness of spin-charge separation phenomena. When a weak screening is added, we observe that the spectral function remains unchanged at positive frequency. On the contrary, the spectral function is affected by the screening at negative frequency. However, only the power laws associated to its singularities are modified, not the positions of the singularities (see the dots in Fig.~\ref{figureSPEC2}). This can be explained by the fact that the power laws of the spin excitations and the charge excitations depend both on the Coulomb parameter in the charge sector\cite{voit2}, and that the charge sector is precisely the one which is affected by the screening potential (see Eq.~(\ref{lagrangian_Sc})).

For weak Coulomb interactions, the numerical convergence starts to become problematic, especially for $K_c>1/2$ which correspond to the most relevant values in our approach (see Sec.~III). For this reason, we decide to study, in the next sections, the tunnel and the local density of states for which numerical convergence is easier to obtain. A more precise study of the spectral function profile is beyond the scope of this paper.

\subsection{Tunnel density of states}

We now turn our interest to the tunnel density of states (TDOS), i.e. the density of states of the wire at $x=0$, which expression in the presence of screening is:

\begin{eqnarray}\label{TDOS}
\rho_{r,\sigma}(0,\Omega)&=&\frac{1}{\pi^2 a} \int_{0}^{+\infty}d\tilde{t}\cos(\Omega \tilde{t})\nonumber\\
&&\times\re\Big[e^{-\nu\ln(1+i\omega_c\tilde{t})+\frac{K_c}{8}f_{Sc}(\tilde{t})}\Big]~,
\end{eqnarray}

where $\nu$ is the non-universal parameter defined as $\nu\equiv 2\gamma_c+1=(K_c+K_c^{-1}+2)/4$, and $f_{Sc}$ corresponds to the contribution due to the screening:
\begin{eqnarray}
&&f_{Sc}(\tilde{t})=-e^{\omega_{Sc}\tilde{t}-i\frac{\omega_{Sc}}{\omega_c}}Ei\left(-\omega_{Sc}\tilde{t}+i\frac{\omega_{Sc}}{\omega_c}\right)\nonumber\\
&&-e^{-\omega_{Sc}\tilde{t}+i\frac{\omega_{Sc}}{\omega_c}}Ei\left(\omega_{Sc}\tilde{t}-i\frac{\omega_{Sc}}{\omega_c}\right)+2\ln(1+i\omega_c\tilde{t})\nonumber\\
&&+e^{-i\frac{\omega_{Sc}}{\omega_c}}Ei\left(i\frac{\omega_{Sc}}{\omega_c}\right)+e^{i\frac{\omega_{Sc}}{\omega_c}}Ei\left(-i\frac{\omega_{Sc}}{\omega_c}\right)~.
\end{eqnarray}

Since we look at the density of states for a non-magnetic system, the final expression of the TDOS does not depend on the spin $\sigma$. Because of the symmetry of the energy bands, it does not depend on the chirality $r$. It is interesting to notice that the logarithm terms in the exponential of Eq.~(\ref{TDOS}) can be rearranged so that the tunnel density of states reads:
\begin{eqnarray}
\rho_{r,\sigma}(0,\Omega)&=&\frac{1}{\pi^2 a} \int_{0}^{+\infty}d\tilde{t}\cos(\Omega \tilde{t})\nonumber\\
&&\times\re\Big[e^{-\tilde{\nu}\ln(1+i\omega_c\tilde{t})+\frac{K_c}{8}\tilde{f}_{Sc}(\tilde{t})}\Big]~,
\end{eqnarray}

with the new exponent $\tilde{\nu}\equiv(K_c^{-1}+2)/4$, and:
\begin{eqnarray}
&&\tilde{f}_{Sc}(\tilde{t})=-e^{\omega_{Sc}\tilde{t}-i\frac{\omega_{Sc}}{\omega_c}}Ei\left(-\omega_{Sc}\tilde{t}+i\frac{\omega_{Sc}}{\omega_c}\right)\nonumber\\
&&-e^{-\omega_{Sc}\tilde{t}+i\frac{\omega_{Sc}}{\omega_c}}Ei\left(\omega_{Sc}\tilde{t}-i\frac{\omega_{Sc}}{\omega_c}\right)\nonumber\\
&&+e^{-i\frac{\omega_{Sc}}{\omega_c}}Ei\left(i\frac{\omega_{Sc}}{\omega_c}\right)+e^{i\frac{\omega_{Sc}}{\omega_c}}Ei\left(-i\frac{\omega_{Sc}}{\omega_c}\right)~.
\end{eqnarray}

\begin{figure}[h!]
\begin{center}
\includegraphics[width=7cm]{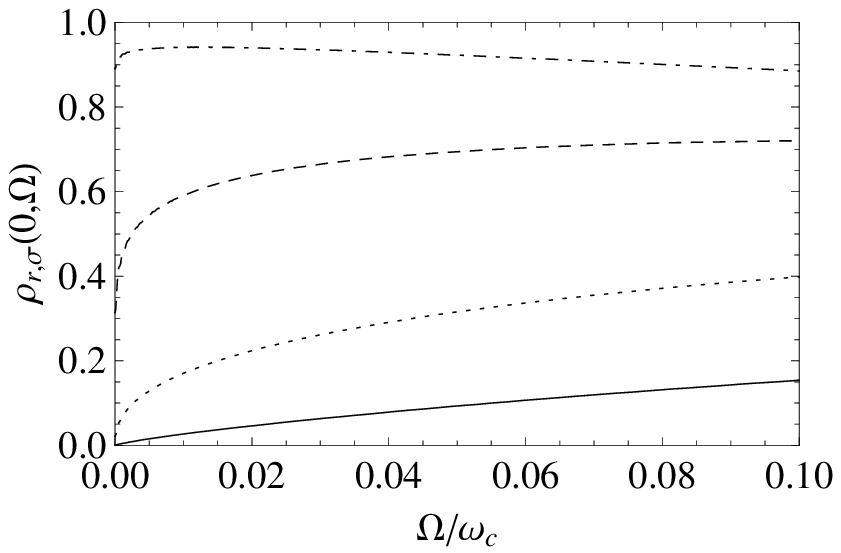}
\includegraphics[width=7cm]{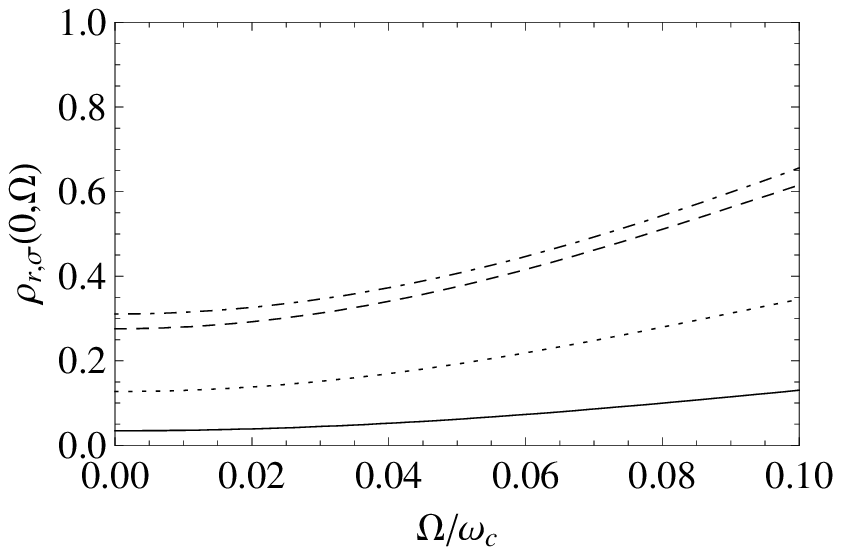}
\caption{Top panel: TDOS as a function of frequency in the absence of screening ($\omega_{Sc}/\omega_c=0$) in unit of $2\pi v_F$. Bottom panel: TDOS in the presence of a strong screening ( $\omega_{Sc}/\omega_c=0.1$). On both graphics, the Coulomb interactions parameter is $K_c=0.2$ (solid line), $K_c=0.3$ (dotted line), $K_c=0.5$ (dashed line), and $K_c=0.8$ (dotted-dashed line).}\label{figureTDOS}
\end{center}
\end{figure}

Thus, in the TDOS, there is an exact compensation of the $-K_c/4$ contribution in the logarithm term due to the screening by the tip regardless of the strength of the screening. Remember that this contribution appears only for the bulk-tunneling geometry\cite{kane}. Indeed, for an end-tunneling geometry, we would have a factor $(K_c^{-1}+1)/2$ whereas for a bulk-tunneling geometry, we have $(K_c+K_c^{-1}+2)/4$ as it is the case here. In summary, the screening suppresses the specific contribution of the bulk-tunneling, in agreement with the result of Ref.~\onlinecite{safi1} where it is shown that the TDOS of a quantum wire probed by a STM corresponds to the end density of states instead of the bulk density of states. Notice however that the result that we obtain here does not correspond exactly to the one that we would have for an end-tunneling because we end up with a factor $\tilde{\nu}=(K_c^{-1}/2+1)/2$ which is different from $(K_c^{-1}+1)/2$.

In Fig.~\ref{figureTDOS}, the TDOS is plotted as a function of the frequency for different values of the Coulomb interactions parameter. Since the TDOS is an even function on frequency (energy), we only represent its variation at positive frequency. The top panel of Fig.~\ref{figureTDOS} corresponds to the TDOS in the absence of screening: it is shown that the behavior of the TDOS depends strongly of the interactions. When the Coulomb interactions are reduced (i.e. when $K_c$ increases), the TDOS becomes almost constant. Indeed, for a non-interacting system the TDOS is constant due to the linear bands at low energy\cite{glazman_fisher}.

The bottom panel of Fig.~\ref{figureTDOS} corresponds to the TDOS in the presence of a strong screening ($\omega_{Sc}/\omega_c=0.1$) when the tip is put very close to the quantum wire (small $d$): indeed, we have $d/a\approx (\sqrt{K_c}/2)/(\omega_{Sc}/\omega_c)$. In that case, at low frequency and for strong Coulomb interactions ($K_c<1/2$), the TDOS is enhanced in comparison to the TDOS in the absence of screening. However, this behavior must be considered with caution since we have shown in Sec.~III that our non perturbative calculations apply only for $K_c>1/2$.

For weak Coulomb interactions ($K_c>1/2$), we observe a reduction of the TDOS and its behavior becomes weakly dependent of the Coulomb interactions parameter (see the dashed and dotted-dashed lines in the bottom panel of Fig.~\ref{figureTDOS}). We conclude that in the regime of weak interactions, the effect of the Coulomb interactions is no more visible due to the effect of the screening by the tip.

\subsection{Local density of states}

For the local density of states (LDOS) at $x\ne 0$, we obtain in the presence of screening:
\begin{eqnarray}\label{LDOS}
&&\rho_{r,\sigma}(x,\Omega)=\frac{1}{\pi^2 a} \int_{0}^{+\infty}d\tilde{t}\cos(\Omega \tilde{t})\nonumber\\
&&\times\re\Big[e^{-\nu\ln(1+i\omega_c\tilde{t})+\frac{K_c^2-1}{16K_c}\sum_{r'}[f_{Sc}^{r'}(x,\tilde{t})-f_{Sc}^{r'}(x,0)]}\Big]~,\nonumber\\
\end{eqnarray}

where
\begin{eqnarray}
&&f_{Sc}^{r'}(x,\tilde{t})=2i\pi r'\cosh\left(\omega_{Sc}\kappa_{r'}\right)+2\ln\left(\kappa_{r'}\right)\nonumber\\
&&-e^{\omega_{Sc}\kappa_{r'}}\ei\left(-\omega_{Sc}\kappa_{r'}\right)
-e^{-\omega_{Sc}\kappa_{r'}}\ei\left(\omega_{Sc}\kappa_{r'}\right)~,
\end{eqnarray}

with $\kappa_{r'}\equiv -r'\tilde{t}-2K_c|x|/v_F+iar'/v_F$.

The LDOS is an even function with the frequency (as the TDOS) and an even function of position $x$ because of the symmetry of the system. The first information we extract from Eq.~(\ref{LDOS}) is that in the absence of Coulomb interactions (i.e. $K_c=1$), the screening part cancels because of the pre-factor $K_c^2-1$ in front of the function $f_{Sc}^{r'}$. It means that in the absence of interactions, the density of states is only affected locally by the screening at position $x=0$ (see Eq.~(\ref{TDOS})). For $x\ne 0$, the density of states stays the same as an isolated non-interacting wire. Alternatively said, it is only in the presence of Coulomb interactions that the density of states is affected by the screening at position $x\ne 0$. 

\begin{figure}[h!]
\begin{center}
\includegraphics[width=7cm]{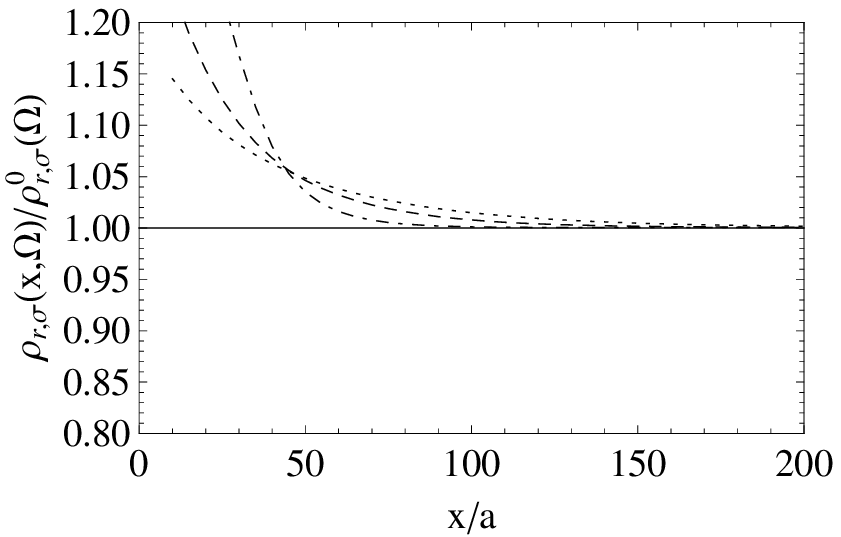}
\includegraphics[width=7cm]{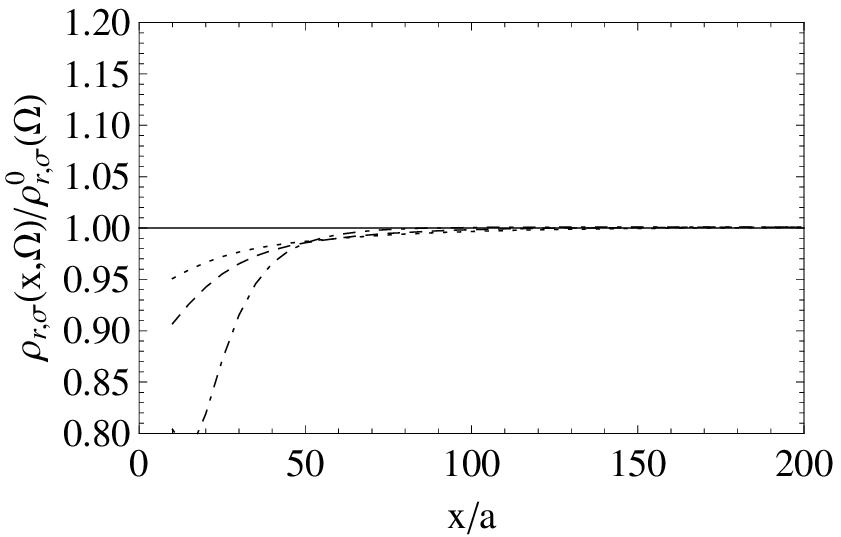}
\caption{Top panel: normalized LDOS as a function of position $x/a$ along the wire for $K_c=0.8$ and  $\Omega/\omega_c=0.01$. Bottom panel: the same quantity at higher frequency: $\Omega/\omega_c=0.05$. On both graphics, the distance between the tip and the wire are $d/a=30$ (dotted line), $d/a=20$ (dashed line), and $d/a=10$ (dotted-dashed line).}\label{figureLDOS}
\end{center}
\end{figure}

In Fig.~\ref{figureLDOS}, the LDOS is plotted as a function of position $x$ along the wire for different values of $d$, where $d$ is the tip-wire distance (see Fig.~\ref{system}). The LDOS is normalized by its unscreened value with is obtained by letting $\omega_{Sc} \to 0$: the difference $f_{Sc}^{r'}(x,\tilde{t})-f_{Sc}^{r'}(x,0)$ cancels in Eq.~(\ref{LDOS}), and we end up with the density of states of a homogeneous interacting one-dimensional system which is position independent:
\begin{eqnarray}
\rho_{r,\sigma}^0(\Omega)&=&\frac{1}{\pi^2 a} \re\left[\int_{0}^{+\infty}d\tilde{t}\frac{\cos(\Omega \tilde{t})}{(1+i\omega_c\tilde{t})^\nu}\right]\nonumber\\
&=&\frac{e^{-|\Omega|/\omega_c}}{2\pi v_F\Gamma(\nu)}\left|\frac{\Omega}{\omega_c}\right|^{\nu-1}~,
\end{eqnarray}

where $\Gamma$ is the Gamma function. On the graphics, we only represent the values of the LDOS for $(x/a)>10$ in order to satisfy the condition $x\gg a\approx\lambda_F$, where $\lambda_F$ is the Fermi wave length. 

We observe a deviation of the LDOS close to the tip position and a convergence toward its unscreened value at large $x$. The top panel of Fig.~\ref{figureLDOS} shows the LDOS for $K_c=0.8$, $\Omega/\omega_c=0.01$ and several tip-wire distance. In this case, the LDOS is increased in comparison to its unscreened value. On the contrary, at higher frequency, the LDOS is below its unscreened value (see the bottom panel of Fig.~\ref{figureLDOS}). The spatial extension of the deviation is related to the Coulomb interactions parameter $K_c$, the frequency $\Omega$ and the tip-wire distance $d$. 

\begin{figure}[ht]
\begin{center}
\includegraphics[width=7cm]{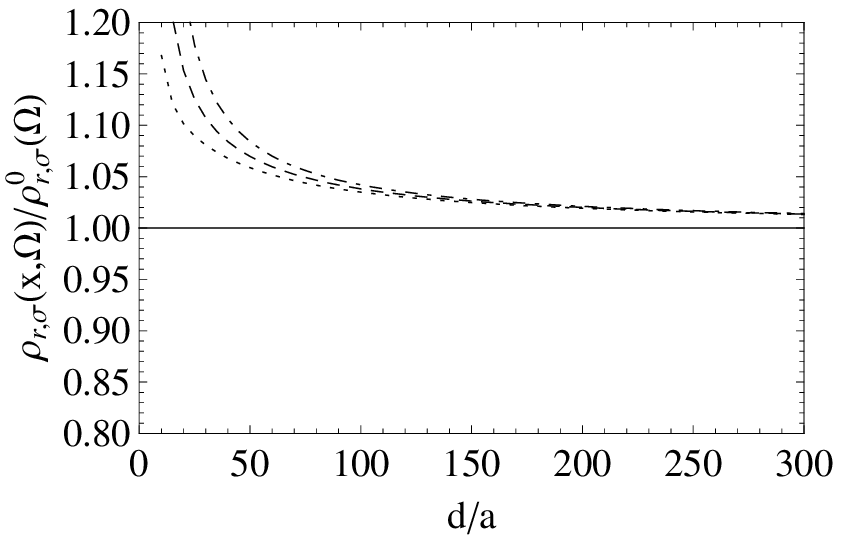}
\includegraphics[width=7cm]{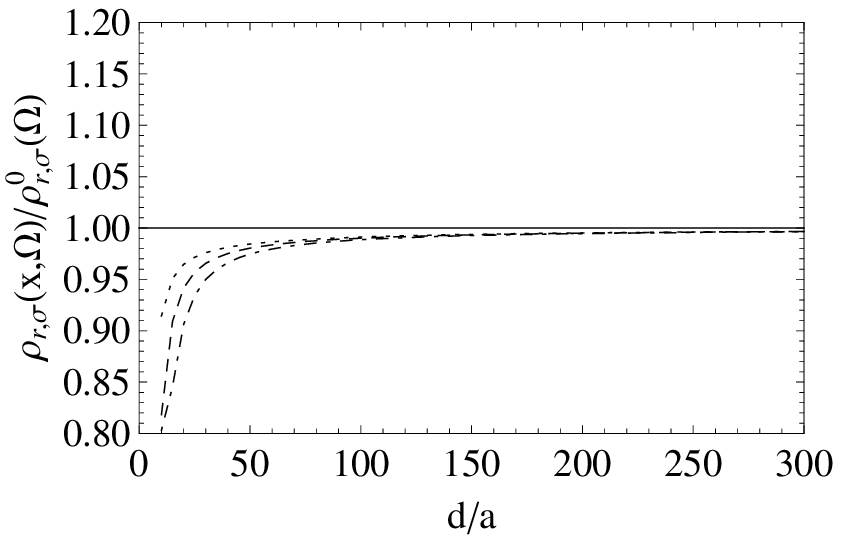}
\caption{Top panel: normalized LDOS as a function of distance $d/a$ between the tip and the wire for $K_c=0.8$ and $\Omega/\omega_c=0.01$. Bottom panel: the same quantity at higher frequency $\Omega/\omega_c=0.05$. On both graphics, the position along the wire are $x/a=30$ (dotted line), $x/a=20$ (dashed line), and $x/a=10$ (dotted-dashed line). }\label{figureDIST}
\end{center}
\end{figure}

In Fig.~\ref{figureDIST}, the LDOS is plotted as a function of the tip-wire distance $d$ for different values of the position $x$. It is shown that the deviation from the unscreened value increases when $d$ decreases, i.e. when the tip gets closer and closer to the wire. In addition, we observe a convergence through the unscreened value of the LDOS when $d$ increases. In a similar way than in Fig.~\ref{figureLDOS}, we obtain larger values of the screened LDOS for $\Omega/\omega_c=0.01$ (top panel of Fig.~\ref{figureDIST}) and smaller values of the screened LDOS for $\Omega/\omega_c=0.05$ (bottom panel of Fig.~\ref{figureDIST}).

In summary, when the STM tip is approached close to the wire, the LDOS shows a deviation from its unscreened value which has a finite spatial extension whose depends both on the frequency and on the Coulomb interactions parameter.


\section{Conclusion}

In this paper, we have studied the electrostatic screening of a quantum wire by a STM tip. We have shown, within reasonable
approximation, how to derive non perturbatively the Dyson equation of the wire in the presence of the tip. Next, we applied our results to the particular case where the double derivative of the potential is local. It allows us to calculate explicitly the Green's function of the wire. The effect of the screening on the spectral properties was then presented, whereas the effect on the transport properties will be presented in a separate paper\cite{guigou}.

The screening by the tip strongly affect the spectral function and TDOS. The power laws associated to the singularities of the spectral function are modified, whereas the positions of these singularities remain unchanged. At low frequency, the TDOS is enhanced for large Coulomb interactions whereas it is reduced for weak Coulomb interactions. In the absence of screening, the LDOS does not depend on the position since we consider a homogeneous wire. On the contrary, in the presence of screening by the tip, the whole system is no more homogeneous and the LDOS becomes strongly dependent of the position along the wire and of the distance between the tip and the wire. However, these dependences are only present when Coulomb interactions are present. This is because we have assumed a local electrostatic potential. That means that, in the case we have considered, the Coulomb interactions is responsible for the spatial extension of the screening. More precisely, we have shown that the spatial extension of the LDOS is related to both the strength of the Coulomb interactions and the strength of the screening potential. 

This work opens the way for a large domain of applications. First, it is possible quite simply to generalize our approach to one-dimensional systems with more degrees of freedom, such as a carbon nanotube for example. Second, it is possible to address the question of electrical reservoirs at the extremities of the wire. In that case, the bare Green's functions are more complicated because they contain reflection terms\cite{safi2,lebedev} at the contacts. However, it should be possible to calculate numerically the full Green's function and to study simultaneously the effect of the screening by the tip and the effect of the reservoirs. Third, with the help of numerical calculations, the study of more realistic potentials, such as the full image charge potential of Eq.~(\ref{potential}), could be envisioned. Finally, the present calculation relies on the 
fact that we have neglected $2k_F$ oscillatory terms in the wire density. Including such terms render the present problem 
not solvable anymore, but such terms could be treated within the context of perturbation theory, which is fully justified because of their irrelevant character.  

\section{acknowledgments}

We thank D.~Bercioux, F.~Dolcini, H.~Grabert, F.~Hekking and I.~Safi for valuable discussions.


\appendix

\section{Dyson equation derivation}

Let start by introducing the partition function as an imaginary time path integral over boson fields\cite{QFT}
\begin{eqnarray}
Z[\varphi,\phi,\theta]&=&\int {\mathcal D}\varphi_{\uparrow}{\mathcal D}\varphi_{\downarrow}{\mathcal D}\phi_c{\mathcal D}\phi_s{\mathcal D}\theta_c{\mathcal D}\theta_s e^{-\int d\tau (L-L_{aux})}~,\nonumber\\
\end{eqnarray}

where $L=L_W+L_T+L_{Sc}$ is the total Lagrangian of the system. From the commutation relation between the fields $\phi_j$ and $\theta_j$, one obtains the expression of the Lagrangian of the quantum wire 
\begin{eqnarray}
L_{W}&=&-\frac{1}{2}\sum_{j} \int_{-\infty}^{+\infty}dx
\left( \begin{array}{cc} 
\phi_j(x,\tau)&\theta_j(x,\tau)
\end{array} \right)\nonumber\\
&&\times G^{-1}_W(x,\tau) 
\left( \begin{array}{c} 
\phi_j(x,\tau) \\ \\
\theta_j(x,\tau) 
\end{array} \right)~,
\end{eqnarray}

where $G_W$ is a $2\times 2$ matrix defined by
\begin{eqnarray}
G^{-1}_W(x,\tau)=\left( \begin{array}{cc} \frac{K_j}{v_j}\partial_{\tau}^2+\partial_xv_jK_j\partial_x & i\partial_x\partial_{\tau}-i\partial_{\tau}\partial_x \\
i\partial_x\partial_{\tau}-i\partial_{\tau}\partial_x & \frac{1}{K_jv_j}\partial_{\tau}^2+\partial_x\frac{v_j}{K_j}\partial_x
\end{array} \right)~.\nonumber\\
\end{eqnarray}

The Lagrangian associated to the STM tip reads\cite{wen}
\begin{eqnarray}
L_{T}=-\frac{1}{2}\sum_{\sigma'} \int_{-\infty}^{+\infty}dy \varphi_{\sigma'}(y,\tau) (G^{\varphi\varphi}_{\sigma'})^{-1}(y,\tau) \varphi_{\sigma'}(y,\tau)~,\nonumber\\
\end{eqnarray}

where $(G^{\varphi\varphi}_{\sigma'})^{-1}(y,\tau)=\frac{1}{2\pi}\partial_y(i\partial_{\tau}+u_F\partial_y)$. The Lagrandian associated to the screening potential is:
\begin{eqnarray}
L_{Sc}=-\sum_{\sigma'} \int_{-\infty}^{+\infty}dx \int_{-\infty}^{+\infty}dy \varphi_{\sigma'}(y,\tau) G^{-1}_{Sc}(x,y)\theta_c(x,\tau)~,\nonumber\\
\label{lagrangian_Sc}
\end{eqnarray}

with $G^{-1}_{Sc}(x,y)=\partial_x\partial_yW(x,y)/(\pi\sqrt{2\pi})$. And finally, a Lagrangian containing auxiliary fields has to be added. Its expression is
\begin{eqnarray}
L_{aux}&=&\sum_{j} \int_{-\infty}^{+\infty}dx  \Big[\eta_{\phi_j}(x,\tau)\phi_j(x,\tau)+\eta_{\theta_j}(x,\tau)\theta_j(x,\tau)\Big]\nonumber\\
&&+\sum_{\sigma'}\int_{-\infty}^{+\infty}dy\eta_{ \varphi_{\sigma'}}(y,\tau)\varphi_{\sigma'}(y,\tau)~.
\end{eqnarray}

We decide first to integrate over the degrees of freedom of the quantum wire. Reporting the relation $G^{-1}_W(x,\tau)G_W(x,\tau,x',\tau')=\delta(x-x')\delta(\tau-\tau')$ in the generating functional formula, one can write 
\begin{widetext}
\begin{eqnarray}
&&Z[\varphi,\phi,\theta]=\int {\mathcal D}\varphi_{\uparrow}{\mathcal D}\varphi_{\downarrow}{\mathcal D}\phi_c{\mathcal D}\phi_s{\mathcal D}\theta_c{\mathcal D}\theta_s~A[\phi,\theta]\exp\Bigg\{-\frac{1}{2}\sum_j\int d\mathbf{x}\int d\mathbf{x}'\nonumber\\
&&\times \left(
\begin{array}{ccc}
\eta_{\phi_j}(\mathbf{x})&&\eta_{\theta_j}(\mathbf{x})+\sum_{\sigma'}\int_{-\infty}^{+\infty} dy\varphi_{\sigma'}(\mathbf{y})G^{-1}_{Sc}(x,y)\delta_{jc}
\end{array}\right)
G_W(\mathbf{x};\mathbf{x}')
\left(
\begin{array}{c}
\eta_{\phi_j}(\mathbf{x}')\\
\\
\eta_{\theta_j}(\mathbf{x}')+\sum_{\sigma'}\int_{-\infty}^{+\infty} dy' \varphi_{\sigma'}(\mathbf{y}')G^{-1}_{Sc}(x',y')\delta_{jc} 
\end{array}\right)\nonumber\\
&&-\sum_{\sigma'}\int d\mathbf{y}\eta_{\varphi_{\sigma'}}(\mathbf{y})\varphi_{\sigma'}(\mathbf{y})-\frac{1}{2}\sum_{\sigma'}\int d\mathbf{y}\varphi_{\sigma'}(\mathbf{y})(G^{\varphi\varphi}_{\sigma'})^{-1}(\mathbf{y})\varphi_{\sigma'}(\mathbf{y})\Bigg\}~,
\end{eqnarray}
\end{widetext}

where, in order to clarify the notations, we introduce a $\delta_{jc}$. Indeed, the spin sector $j=s$ is never affected by the screening effect, as we can see from Eq.~(\ref{lagrangian_Sc}). The new label $\mathbf{x}$ ($\mathbf{y}$) denotes the couple of coordinates $(x,\tau)$ (respectively $(y,\tau)$) and $\int d\mathbf{x}\equiv\int d\tau \int_{-\infty}^{+\infty}dx$ (respectively $\int d\mathbf{y}\equiv\int d\tau \int_{-\infty}^{+\infty}dy$). Then
\begin{widetext}
\begin{eqnarray}
A[\phi,\theta]&=&\exp\Bigg\{\frac{1}{2}\sum_j\int d\mathbf{x} \int d\mathbf{x}'\nonumber\\
&\times&\left[\phi_j(\mathbf{x})-\eta_{\phi_j}(\mathbf{x})G_W(\mathbf{x};\mathbf{x}')~\theta_j(\mathbf{x})-\left(\eta_{\theta_j}(\mathbf{x})+\sum_{\sigma'}\int_{-\infty}^{+\infty}dy \varphi_{\sigma'}(\mathbf{y})G^{-1}_{Sc}(x,y)\delta_{jc}\right)G_W(\mathbf{x};\mathbf{x}')\right]\nonumber\\
&\times& G^{-1}_W(\mathbf{x};\mathbf{x}')
\left( 
\begin{array}{c}
\phi_j(\mathbf{x}')-\eta_{\phi_j}(\mathbf{x}')G_W(\mathbf{x};\mathbf{x}') 
\\
\\ \theta_j(\mathbf{x}')-(\eta_{\theta_j}(\mathbf{x}')+\sum_{\sigma'}\int_{-\infty}^{+\infty} dy' \varphi_{\sigma'}(\mathbf{y}')G^{-1}_{Sc}(x',y')\delta_{jc})G_W(\mathbf{x};\mathbf{x}')
\end{array} \right)\Bigg\}~.
\end{eqnarray}
\end{widetext}

Then, we regroup all the terms which are linear in $\varphi_{\sigma'}$. This manipulation helps to highlight the inverse of the propagator $G_{\sigma'}^{\varphi\varphi}$ in presence of screening effects. So, one obtains
\begin{widetext}
\begin{eqnarray}
Z[\varphi,\phi,\theta]&=&\int {\mathcal D}\varphi_{\uparrow}{\mathcal D}\varphi_{\downarrow}{\mathcal D}\phi_c{\mathcal D}\phi_s{\mathcal D}\theta_c{\mathcal D}\theta_s~A[\phi,\theta]
\exp\Bigg\{- \frac{1}{2}\sum_j \Big[\int d\mathbf{x}\int d\mathbf{x}'(\eta_{\phi_j}(\mathbf{x})~\eta_{\theta_j}(\mathbf{x}))G_W(\mathbf{x};\mathbf{x}')\left( \begin{array}{c} \eta_{\phi_j}(\mathbf{x}') \\ \\ \eta_{\theta_j}(\mathbf{x}') \end{array} \right)\nonumber\\
&&-2\sum_{\sigma'}\int d\mathbf{y}\eta_{\varphi_{\sigma'}}(\mathbf{y})\varphi_{\sigma'}(\mathbf{y})-\sum_{\sigma'}\int d\mathbf{y}\int d\mathbf{y}'\varphi_{\sigma'}(\mathbf{y})(\mathbf{G}^{\varphi\varphi}_{\sigma'})^{-1}(\mathbf{y};\mathbf{y}')\varphi_{\sigma'}(\mathbf{y}')\nonumber\\
&&+\sum_{\sigma',j}\delta_{jc}\int d\mathbf{x}\int d\mathbf{x}'\int_{-\infty}^{+\infty}dy'\left[\eta_{\phi_j}(\mathbf{x})G_j^{\phi\theta}(\mathbf{x};\mathbf{x}')+\eta_{\theta_j}(\mathbf{x})G_j^{\theta\theta}(\mathbf{x};\mathbf{x}')\right]\varphi_{\sigma'}(\mathbf{y}')G^{-1}_{Sc}(x',y')\nonumber\\
&&+\sum_{\sigma',j}\delta_{jc}\int d\mathbf{x}\int d\mathbf{x}'\int_{-\infty}^{+\infty}dy\varphi_{\sigma'}(\mathbf{y})G^{-1}_{Sc}(x,y)\left[G_j^{\theta\phi}(\mathbf{x};\mathbf{x}')\eta_{\phi_j}(\mathbf{x}')+G_j^{\theta\theta}(\mathbf{x};\mathbf{x}')\eta_{\theta_j}(\mathbf{x}')\right]\Big]\Bigg\}~,\nonumber\\
\label{gen_func}
\end{eqnarray}
\end{widetext}

where
\begin{eqnarray}
&&(\mathbf{G}^{\varphi\varphi}_{\sigma'})^{-1}(\mathbf{y};\mathbf{y}')=(G^{\varphi\varphi}_{\sigma'})^{-1}(\mathbf{y};\mathbf{y}')\delta(y-y')\delta(\tau-\tau')\nonumber\\
&-&\delta_{jc}\int_{-\infty}^{+\infty}d\mathbf{x}\int_{-\infty}^{+\infty}d\mathbf{x}'G^{-1}_{Sc}(x,y)G_j^{\theta\theta}(\mathbf{x};\mathbf{x}')G^{-1}_{Sc}(x',y')~.\nonumber\\
\end{eqnarray}

Since the elements of the matrix $G_W$ are developed, we use the properties of bosonic Green functions in the Matsubara formalism: $G^{\phi\theta}_j(\mathbf{x};\mathbf{x}')=G^{\theta\phi}_j(\mathbf{x}';\mathbf{x})$ and
$G^{\theta\theta}_j(\mathbf{x};\mathbf{x}')=G^{\theta\theta}_j(\mathbf{x}';\mathbf{x})$,
and introducing the relation 
\begin{eqnarray}
(\mathbf{G}^{\varphi\varphi}_{\sigma'})^{-1}(\mathbf{y};\mathbf{y}')\mathbf{G}^{\varphi\varphi}_{\sigma'}(\mathbf{y};\mathbf{y}')&=&\mathbf{G}^{\varphi\varphi}_{\sigma'}(\mathbf{y};\mathbf{y}')(\mathbf{G}^{\varphi\varphi}_{\sigma'})^{-1}(\mathbf{y};\mathbf{y}')\nonumber\\
&=&\delta(y-y')\delta(\tau-\tau')~, 
\end{eqnarray}

the generating functional is expressed by
\begin{widetext}
\begin{eqnarray}
Z[\varphi,\phi,\theta]&=&\int {\mathcal D}\varphi_{\uparrow}{\mathcal D}\varphi_{\downarrow}{\mathcal D}\phi_c{\mathcal D}\phi_s{\mathcal D}\theta_c{\mathcal D}\theta_s~A[\phi,\theta]B[\varphi]
\exp\{-\frac{1}{2}\sum_j\int d\mathbf{x}\int d\mathbf{x}'(\eta_{\phi_j}(\mathbf{x})~\eta_{\theta_j}(\mathbf{x}))G_W(\mathbf{x};\mathbf{x}')\left( \begin{array}{c} \eta_{\phi_j}(\mathbf{x}') \\ \\ \eta_{\theta_j}(\mathbf{x}') \end{array} \right)\nonumber\\
&+&\frac{1}{2}\delta_{jc}\sum_{\sigma',j}\int d\mathbf{y}\int d\mathbf{y}'\int d\mathbf{x_1} \int d\mathbf{x'_2}\int_{-\infty}^{+\infty}dx_1'\int_{-\infty}^{+\infty}dx_2\nonumber\\
&\times&\Big[\eta_{\phi_j}(\mathbf{x_1})G_j^{\phi\theta}(\mathbf{x_1};x_1',\tau)G^{-1}_{Sc}(x_1',y')+\eta_{\theta_j}(\mathbf{x_1})G_j^{\theta\theta}(\mathbf{x_1};x_1',\tau)G^{-1}_{Sc}(x_1',y')-\eta_{\varphi_{\sigma'}}(\mathbf{y})\Big
](\mathbf{G}^{\varphi\varphi}_{\sigma'})^{-1}(\mathbf{y};\mathbf{y}')\nonumber\\
&\times&\Big[G^{-1}_{Sc}(x_2,y')G_j^{\phi\theta}(\mathbf{x_2'};x_2,\tau')\eta_{\phi_j}(\mathbf{x_2'})+G^{-1}_{Sc}(x_2,y')G_j^{\theta\theta}(\mathbf{x_2'};x_2,\tau')\eta_{\theta_j}(\mathbf{x_2'})-\eta_{\varphi_{\sigma'}}(\mathbf{y}')\Big]\}~,
\label{fonc_part}
\end{eqnarray}

with
\begin{eqnarray}
B[\varphi]&=&\exp\Bigg\{\frac{1}{2}\sum_{\sigma'}\int d\mathbf{y} \int d\mathbf{y}'\Big[\Big(\varphi_{\sigma'}(\mathbf{y})+(-M_1(\mathbf{y})-M_2(\mathbf{y})+\eta_{\varphi_{\sigma'}}(\mathbf{y}))\mathbf{G}^{\varphi\varphi}_{\sigma'}(\mathbf{y};\mathbf{y}')\Big)\nonumber\\
&\times&(\mathbf{G}^{\varphi\varphi}_{\sigma'})^{-1}(\mathbf{y};\mathbf{y}')\Big(\varphi_{\sigma'}(\mathbf{y}')+\mathbf{G}^{\varphi\varphi}_{\sigma'}(\mathbf{y};\mathbf{y}')(-M_1(\mathbf{y})-M_2(\mathbf{y})+\eta_{\varphi_{\sigma'}}(\mathbf{y}))\Big)\nonumber\\
&-&(-M_1(\mathbf{y})-M_2(\mathbf{y})+\eta_{\varphi_{\sigma'}}(\mathbf{y}))\mathbf{G}^{\varphi\varphi}_{\sigma'}(\mathbf{y};\mathbf{y}')(-M_1(\mathbf{y}')-M_2(\mathbf{y}')+\eta_{\varphi_{\sigma'}}(\mathbf{y}'))\Big]\Bigg\}~,
\end{eqnarray}
\end{widetext}

where $M_1(\mathbf{y})$ (respectively $M_2(\mathbf{y})$) contains mixed terms of $\eta_{\phi_j}$ and $\varphi_{\sigma'}$ (respectively $\eta_{\theta_j}$ and $\varphi_{\sigma'}$) in Eq.~(\ref{gen_func}).

From Eq.~(\ref{fonc_part}), one can obtain any correlation functions, using 
\begin{eqnarray}
\langle\theta_j(\mathbf{x})\theta_j(\mathbf{x}')\rangle&=&\mathbf{G}_j^{\theta\theta}(\mathbf{x};\mathbf{x}')=\frac{1}{Z}\frac{\partial^2Z}{\partial\eta_{\theta_j}(\mathbf{x})\partial\eta_{\theta_j}(\mathbf{x}')}~.\nonumber\\
\label{fct_green}
\end{eqnarray}

Finally, the Dyson equation in the Matsubara formalism for the bosonic Green's function associated to the field $\theta_j$ reads:
\begin{widetext}
\begin{eqnarray}\label{dyson2}
\mathbf{G}^{\theta\theta}_j(x,\tau;x'\tau')&=&G^{\theta\theta}_j(x,\tau;x'\tau')+\delta_{jc}\sum_{\sigma'}\int d\tau_1\int d\tau_2\int_{-\infty}^{+\infty}dx_1\int_{-\infty}^{+\infty}dx_2\int_{-\infty}^{+\infty}dy_1\int_{-\infty}^{+\infty}dy_2\nonumber\\
&\times& G^{\theta\theta}_j(x,\tau;x_1,\tau_1)G^{-1}_{Sc}(x_1,y_1)\mathbf{G}^{\varphi\varphi}_{\sigma'}(y_1,\tau_1;y_2,\tau_2)G^{-1}_{Sc}(x_2,y_2)G^{\theta\theta}_j(x_2,\tau_2;x',\tau')~.
\end{eqnarray}
\end{widetext}

The derivation of the partition function with other auxiliary fields gives access, in the same way, to the other correlators. Indeed, for example
\begin{eqnarray}
&&\mathbf{G}^{\varphi\theta}_{j\sigma'}(y,\tau;x',\tau')=-\delta_{jc}\int d\tau_1\int_{-\infty}^{+\infty}dx_1\int_{-\infty}^{+\infty}dy_1\nonumber\\
&&\times \mathbf{G}^{\varphi\varphi}_{\sigma'}(y,\tau;y_1,\tau_1)G^{-1}_{Sc}(x_1,y_1)G^{\theta\theta}_j(x_1,\tau_1;x',\tau')~.\nonumber\\
\end{eqnarray}

In order to find the Dyson equation of the Green's function of the STM tip, we start by integrating out the degrees of freedom of the tip. From the same partition function, one finds:
\begin{widetext}
\begin{eqnarray}
\mathbf{G}^{\varphi\varphi}_{\sigma'}(y,\tau;y',\tau')&=&G^{\varphi\varphi}_{\sigma'}(y,\tau;y',\tau')+\sum_{j}\delta_{jc}\int d\tau_1 \int d\tau_2\int_{-\infty}^{+\infty}dx_1\int_{-\infty}^{+\infty}dx_2\int_{-\infty}^{+ \infty}dy_1\int_{-\infty}^{+\infty}dy_2\nonumber\\
&\times&G^{\varphi\varphi}_{\sigma'}(y,\tau;y_1,\tau_1)G^{-1}_{Sc}(x_1,y_1)\mathbf{G}_j^{\theta\theta}(x_1,\tau_1;x_2,\tau_2)G^{-1}_{Sc}(x_2,y_2)G^{\varphi\varphi}_{\sigma'}(y_2,\tau_2;y',\tau')~.
\end{eqnarray}
\end{widetext}


\section{Expressions of the bare Green's functions}

For an infinite wire length and in the absence of screening, the Matsubara bosonic Green's functions at zero temperature are:
\begin{eqnarray}
G^{\phi\phi}_{c}(x,x',\omega)&=&\frac{1}{2|\omega|K_{c}}e^{-|\omega| |x-x'|/v_c}~,\\
G^{\theta\theta}_{c}(x,x',\omega)&=&\frac{K_{c}}{2|\omega|}e^{-|\omega| |x-x'|/v_c}~,\\
G^{\theta\phi}_{c}(x,x',\omega)&=&G^{\phi\theta,0}_{c}(x,x',\omega)\nonumber\\
&=&\frac{sgn(x-x')}{2\omega}e^{-|\omega||x-x'|/v_c}~.
\end{eqnarray}

The bosonic Green's function of the tip which we need in the calculation is the one located at $y=y'=0$ which is given by the expression at zero temperature:
\begin{eqnarray}
G_{\sigma'}^{\varphi\varphi}(0,0,\omega)&=&\frac{\pi}{|\omega|}~.
\end{eqnarray}


\section{Expressions of the Green's functions in the presence of screening}

By iteration, one can easily show that the Dyson equation given by Eq.~(\ref{dyson2}) for $j=c$ can be re-written as
\begin{widetext}
\begin{eqnarray}
\mathbf{G}^{\theta\theta}_c(x,\tau;x'\tau')&=&G^{\theta\theta}_c(x,\tau;x'\tau')+\sum_{\sigma'}\int d\tau_1\int d\tau_2\int_{-\infty}^{+\infty}dx_1\int_{-\infty}^{+\infty}dx_2\int_{-\infty}^{+\infty}dy_1\int_{-\infty}^{+\infty}dy_2\nonumber\\
&\times& G^{\theta\theta}_c(x,\tau;x_1,\tau_1)G_{Sc}^{-1}(x_1,y_1)G_{\sigma'}^{\varphi\varphi}(y_1,\tau_1;y_2,\tau_2)G_{Sc}^{-1}(x_2,y_2)\mathbf{G}^{\theta\theta}_c(x_2,\tau_2;x',\tau')~.
\end{eqnarray}
\end{widetext}

Let us remind the expression of the operator $G^{-1}_{Sc}$ when we assume a local double derivative potential
\begin{eqnarray}
G^{-1}_{Sc}(x,y)&=&\frac{1}{\pi \sqrt{2\pi}}\partial_{x}\partial_{y}W(x,y)
=\frac{W_0}{\pi \sqrt{2\pi}}\delta(y)\delta(x)~.\nonumber\\
\end{eqnarray}

Next, using the time translation invariance, one makes the Fourier transform of the Dyson equation and one obtains
\begin{eqnarray}\label{C3}
&&\mathbf{G}^{\theta\theta}_c(x,x',\omega)=G^{\theta\theta}_c(x,x',\omega)\nonumber\\
&&+\sum_{\sigma'}\frac{W_0^2}{2\pi^3}G^{\theta\theta}_c(x,0,\omega)G_{\sigma'}^{\varphi\varphi}(0,0,\omega)\mathbf{G}^{\theta\theta}_c(0,x',\omega)~.\nonumber\\
\end{eqnarray}

The bare Matsubara Green functions of the tip and the wire are already known in frequency (see Appendix B). Thus, one replaces their expressions in Eq.~(\ref{C3}) and one performs the sum over $\sigma'$
\begin{eqnarray}
\mathbf{G}^{\theta\theta}_{c}(x,x',\omega)&=&G^{\theta\theta}_{c}(x,x',\omega)\nonumber\\
&+&\frac{ K_c\omega_{Sc}^2}{2|\omega|(\omega^2-\omega_{Sc}^2)}e^{-|\omega|K_c\frac{|x|+|x'|}{v_F}}~,
\end{eqnarray}

where $\omega_{Sc}^2\equiv K_cW_0^2/(2\pi^2)$. After performing an inverse Fourier transform, one finds for the second term which corresponds to the screening part
\begin{eqnarray}\label{eq_integrale}
\mathbf{G}^{\theta\theta}_{c,Sc}(x,x',\tilde{\tau})&=&\frac{K_c\omega_{Sc}^2}{4\pi}\sum_r\int_{0}^{+\infty}d\omega \frac{e^{\omega\left(ir\tilde{\tau}-K_c\frac{|x|+|x'|}{v_F}\right)}}{\omega(\omega^2-\omega_{Sc}^2)}~.\nonumber\\
\end{eqnarray}

The function which appears in Eq.~(\ref{eq_integrale}) can be integrated:
\begin{eqnarray}
&&\int d\omega\frac{e^{\omega Z_r}}{\omega(\omega^2-\omega_{Sc}^2)}=-\ei\left(\omega Z_r\right)\nonumber\\
&&
+\frac{1}{2}\Big[e^{\omega_{Sc}Z_r}\ei\left((\omega-\omega_{Sc})Z_r\right)+e^{-\omega_{Sc}Z_r}\ei\left((\omega+\omega_{Sc})Z_r\right)\Big]~,\nonumber\\
\end{eqnarray}

where $\ei$ is the exponential integral function, and $Z_r\equiv ir\tilde{\tau}-K_c(|x|+|x'|)/v_F$.

The transition between Matsubara and Keldysh formalisms is obtained by an analytic continuation over the time $\tilde{\tau}~\rightarrow~i\tilde{t}+\tau_0$ where $\tau_0=a/v_F$. It allows to retrieve the screening part of the correlator $\mathbf{G}^{\theta\theta,-+}_{c}$:
\begin{eqnarray}
&&\mathbf{G}^{\theta\theta,-+}_{c,Sc}(x,x',\tilde{t})=\frac{K_c}{8\pi}\sum_r\Bigg[2i\pi r\cosh\left(\omega_{Sc}\chi _r\right)\nonumber\\
&&+2\ln\left(\chi _r\right)-e^{\omega_{Sc}\chi _r}\ei\left(-\omega_{Sc}\chi _r\right)
-e^{-\omega_{Sc}\chi _r}\ei\left(\omega_{Sc}\chi _r\right)\Bigg]~,\nonumber\\
\end{eqnarray}

where $\chi _r\equiv -r\tilde{t}-K_c(|x|+|x'|)/v_F+iar/v_F$ and where we have subtracted the terms $\mathbf{G}^{\theta\theta,-+}_{c,Sc}(x,x,0)/2$ and $\mathbf{G}^{\theta\theta,-+}_{c,Sc}(x',x',0)/2$ because of the definition of Eq.~(\ref{Definition_Green}). The four Keldysh Green's functions, which correspond to the four possible arrangements between the indexes $\eta$ and $\eta'$ of the two branches on the Keldysh contour, are the elements of the $2\times 2$ matrix
\begin{eqnarray}
&&\mathbf{G}_{c(K)}^{\theta\theta}(x,x',\tilde{t})=\nonumber\\
&&\left (
\begin{array}{ccc}
\begin{array}{ll} \tilde{t}>0~: &\mathbf{G}_{c}^{\theta\theta}(x,x',\tilde{t})\\
\tilde{t}<0~: & \mathbf{G}_{c}^{\theta\theta}(x',x,-\tilde{t}) \end{array}
 & \mathbf{G}_{c}^{\theta\theta}(x',x,-\tilde{t})\\
\mathbf{G}_{c}^{\theta\theta}(x,x',\tilde{t}) & \begin{array}{ll} \tilde{t}>0~: &\mathbf{G}_{c}^{\theta\theta}(x',x,-\tilde{t})\\
\tilde{t}<0~: & \mathbf{G}_{c}^{\theta\theta}(x,x',\tilde{t}) \end{array}
\end{array} \right )~.\nonumber\\
\end{eqnarray}

In a similar way, we calculate the correlator of the other bosonic fields, we obtain:
\begin{eqnarray}
&&\mathbf{G}^{\phi\phi,-+}_{c,Sc}(x,x',\tilde{t})=\frac{\sgn(x)\sgn(-x')}{8\pi K_c}\nonumber\\
&&\times\sum_r\Bigg[2i\pi r\cosh\left(\omega_{Sc}\chi _r\right)+2\ln\left(\chi _r\right)\nonumber\\
&&-e^{\omega_{Sc}\chi _r}\ei\left(-\omega_{Sc}\chi _r\right)
-e^{-\omega_{Sc}\chi _r}\ei\left(\omega_{Sc}\chi _r\right)\Bigg]~,
\end{eqnarray}

and,
\begin{eqnarray}
&&\mathbf{G}^{\phi\theta,-+}_{c,Sc}(x,x',\tilde{t})=\frac{\sgn(x)}{8\pi}\nonumber\\
&&\times\sum_r r\Bigg[2i\pi r\cosh\left(\omega_{Sc}\chi _r\right)+2\ln\left(\chi _r\right)\nonumber\\
&&-e^{\omega_{Sc}\chi _r}\ei\left(-\omega_{Sc}\chi _r\right)
-e^{-\omega_{Sc}\chi _r}\ei\left(\omega_{Sc}\chi _r\right)\Bigg]~,\nonumber\\
\end{eqnarray}

and finally,
\begin{eqnarray}
&&\mathbf{G}^{\theta\phi,-+}_{c,Sc}(x,x',\tilde{t})=\frac{\sgn(-x')}{8\pi}\nonumber\\
&&\times\sum_r r\Bigg[2i\pi r\cosh\left(\omega_{Sc}\chi _r\right)+2\ln\left(\chi _r\right)\nonumber\\
&&-e^{\omega_{Sc}\chi _r}\ei\left(-\omega_{Sc}\chi _r\right)
-e^{-\omega_{Sc}\chi _r}\ei\left(\omega_{Sc}\chi _r\right)\Bigg]~.\nonumber\\
\end{eqnarray}

From Eq.~(\ref{dyson3}), the Dyson equation of the Green's function of the tip, in frequency, reads
\begin{eqnarray}
&&\mathbf{G}_{\sigma'}^{\varphi\varphi}(0,0,\omega)=G_{\sigma'}^{\varphi\varphi}(0,0,\omega)\nonumber\\
&&+\frac{W_0^2}{\pi^3 }G_{\sigma'}^{\varphi\varphi}(0,0,\omega)G^{\theta\theta}_{c}(0,0,\omega)\mathbf{G}_{\sigma'}^{\varphi\varphi}(0,0,\omega)~,\nonumber\\
\end{eqnarray}

where we have assumed to have a non-magnetic tip. The solution is
\begin{eqnarray}
\mathbf{G}_{\sigma'}^{\varphi\varphi}(0,0,\omega)&=&G_{\sigma'}^{\varphi\varphi}(0,0,\omega)+\frac{\pi\omega_{Sc}^2}{|\omega| (\omega^2-\omega_{Sc}^2)}~.\nonumber\\
\end{eqnarray}

Its inverse Fourier transform leads to an expression similar to the one found for bosonic Green's function of the quantum wire. So that, upon the calculation of the integral over frequencies, the passage from Matsubara to Keldysh formalism allows to find the expression of the screening part of the STM tip Green's function, which is
\begin{eqnarray}
&&\mathbf{G}^{\varphi\varphi,-+}_{\sigma',Sc}(0,0,\tilde{t})=\frac{1}{4}\sum_r\Bigg[2\ln\left(-r\tilde{t}+ir\frac{a}{v_F}\right)\nonumber\\
&&-e^{\omega_{Sc}\left(\tilde{t}-i\frac{a}{v_F}\right)}\ei\left(-\omega_{Sc}\left(\tilde{t}-i\frac{a}{v_F}\right)\right)\nonumber\\
&&-e^{-\omega_{Sc}\left(\tilde{t}-i\frac{a}{v_F}\right)}\ei\left(\omega_{Sc}\left(\tilde{t}-i\frac{a}{v_F}\right)\right)\Bigg]~.
\end{eqnarray}


\end{document}